\documentclass[acmsmall]{acmart}

\usepackage{multirow}
\usepackage{multicol}
\usepackage{tabularx}
\usepackage{wrapfig}
\usepackage{threeparttable}
\usepackage{enumitem}
\usepackage{bbding}
\usepackage{subcaption}
\usepackage{pifont}
\usepackage{utfsym}
\usepackage{wasysym}
\usepackage{makecell}
\usepackage[most]{tcolorbox}

\definecolor{MidnightBlue}{HTML}{006895}

\newcommand{\code}[1]{{\fontfamily{cmtt}\fontseries{m}\fontshape{n}\selectfont\small{#1}}}

\makeatletter
\tcbset{
    myhbox/.style 2 args={%
        enhanced,
        breakable,
        left=1pt,right=1pt,top=1pt,bottom=1pt,boxrule=0.7pt,
        drop fuzzy shadow,
        attach boxed title to top left={yshift*=-\tcboxedtitleheight},
        title={#2},
        boxed title size=title,
        boxed title style={%
            sharp corners,
            rounded corners=northwest,
            colback=tcbcolframe,
            boxrule=0pt,
        },
        underlay boxed title={%
            \path[fill=tcbcolframe] (title.south west)--(title.south east)
                to[out=0, in=180] ([xshift=5mm]title.east)--
                (title.center-|frame.east)
                [rounded corners=\kvtcb@arc] |-
                (frame.north) -| cycle;
        },
        #1
    },
    myvbox/.style 2 args={%
        enhanced,
        colback=white,
        colframe=blue!30!black,
        left=8mm,
        overlay={
            \node[rotate=90, anchor=north west, inner sep=2mm, text=white] (title@aux) at (frame.south west) {#2};
            \path[fill=tcbcolframe] (title@aux.south west)--(title@aux.south east)
                to[out=90, in=270] ([yshift=5mm]title@aux.east)--
                (title@aux.center|-frame.north)
                [rounded corners=\kvtcb@arc] -|
                (frame.west) |- (title@aux.west)[sharp corners] -- cycle;
            \node[rotate=90, inner sep=2mm, text=white] at (title@aux) {#2};
        },
        #1
    },
}
\makeatother

\newtcolorbox{myhbox}[2][]{%
    myhbox={#1}{#2}
}

\AtBeginDocument{%
  \providecommand\BibTeX{{%
    \normalfont B\kern-0.5em{\scshape i\kern-0.25em b}\kern-0.8em\TeX}}}

\setcopyright{acmcopyright}
\copyrightyear{2025}
\acmYear{2025}
\acmDOI{XXXXXXX.XXXXXXX}

\begin{document}

\title{A Systematic Literature Review on Explainability for Machine/Deep Learning-based Software Engineering Research}

\author{Sicong Cao}
%\orcid{0000-0003-3688-4437}
\email{DX120210088@yzu.edu.cn}
\affiliation{%
  \department{School of Information Engineering}
  \institution{Yangzhou University}
  \city{Yangzhou}
  \country{China}}
  
\author{Xiaobing Sun}
%\orcid{0000-0001-5165-5080}
\authornote{Corresponding author}
\email{xbsun@yzu.edu.cn}
\affiliation{%
  \department{School of Information Engineering}
  \institution{Yangzhou University}
  \city{Yangzhou}
  \country{China}}

\author{Ratnadira Widyasari}
%\orcid{0000-0001-8190-5458}
\email{ratnadiraw.2020@phdcs.smu.edu.sg}
\affiliation{%
  \department{School of Computing and Information Systems}
  \institution{Singapore Management University}
  \city{Singapore}
  \country{Singapore}}
  
\author{David Lo}
%\orcid{0000-0002-4367-7201}
\email{davidlo@smu.edu.sg}

\affiliation{%
  \department{School of Computing and Information Systems}
  \institution{Singapore Management University}
  \city{Singapore}
  \country{Singapore}}

\author{Xiaoxue Wu}
%\orcid{0009-0009-5432-651X}
\email{xiaoxuewu@yzu.edu.cn}
\affiliation{%
  \department{School of Information Engineering}
  \institution{Yangzhou University}
  \city{Yangzhou}
  \country{China}}

\author{Lili Bo}
%\orcid{0000-0002-7267-4923}
\email{lilibo@yzu.edu.cn}
\affiliation{%
  \department{School of Information Engineering}
  \institution{Yangzhou University}
  \city{Yangzhou}
  \country{China}}

\author{Jiale Zhang}
%\orcid{0000-0002-2143-5666}
\email{jialezhang@yzu.edu.cn}
\affiliation{%
  \department{School of Information Engineering}
  \institution{Yangzhou University}
  \city{Yangzhou}
  \country{China}}

\author{Bin Li}
%\orcid{0000-0001-8500-9917}
\email{lb@yzu.edu.cn}
\affiliation{%
  \department{School of Information Engineering}
  \institution{Yangzhou University}
  \city{Yangzhou}
  \country{China}}

\author{Wei Liu}
%\orcid{0000-0001-8503-4063}
\email{weiliu@yzu.edu.cn}

\affiliation{%
  \department{School of Information Engineering}
  \institution{Yangzhou University}
  \city{Yangzhou}
  \country{China}}

\author{Di Wu}
%\orcid{0000-0002-4753-8161}
\affiliation{%
  \department{School of Mathematics, Physics, and Computing}
  \institution{University of Southern Queensland}
  \city{Toowoomba}
  \country{Australia}}
\email{di.wu@unisq.edu.au}

\author{Yixin Chen}
%\orcid{0000-0002-3704-4432}
\affiliation{%
  \department{Department of Computer Science and Engineering}
  \institution{Washington University in St. Louis}
  \city{St. Louis}
  \country{USA}}
\email{chen@cse.wustl.edu}

\renewcommand{\shortauthors}{S. Cao et al.}
\renewcommand{\shorttitle}{A Systematic Literature Review on Explainability for ML/DL-based SE Research}

\begin{abstract}
The remarkable achievements of Artificial Intelligence (AI) algorithms, particularly in Machine Learning (ML) and Deep Learning (DL), have fueled their extensive deployment across multiple sectors, including Software Engineering (SE). However, due to their black-box nature, these promising AI-driven SE models are still far from being deployed in practice. This lack of explainability poses unwanted risks for their applications in critical tasks, such as vulnerability detection, where decision-making transparency is of paramount importance. This paper endeavors to elucidate this interdisciplinary domain by presenting a systematic literature review of approaches that aim to improve the explainability of AI models within the context of SE. The review canvasses work appearing in the most prominent SE \& AI conferences and journals, and spans 108 papers across 23 unique SE tasks. Based on three key Research Questions (RQs), we aim to (1) summarize the SE tasks where XAI techniques have shown success to date; (2) classify and analyze different XAI techniques; and (3) investigate existing evaluation approaches. Based on our findings, we identified a set of challenges remaining to be addressed in existing studies, together with a set of guidelines highlighting potential opportunities we deemed appropriate and important for future work.

\end{abstract}





\maketitle

\section{Introduction}
\textbf{Software Engineering (SE)} is a discipline that deals with the design, development, testing, and maintenance of software systems. As software continues to pervade a wide range of industries, diverse and complex SE data, such as source code, bug reports, and test cases, have grown to become unprecedentedly large and complex. Driven by the success of \textbf{Artificial Intelligence (AI)} algorithms in various research fields, the SE community has shown great enthusiasm for exploring and applying advanced \textbf{Machine Learning (ML)}/\textbf{Deep Learning (DL)} models to automate or enhance SE tasks typically performed manually by developers, including automated program repair \cite{DBLP:conf/qrs/TangLBWCS21,DBLP:conf/icse/JiangL021}, code generation \cite{DBLP:journals/tse/LiuSZNLZ22}, and vulnerability detection \cite{BGNN4VD,Devign}. A recent report from the 2021 SEI Educator's Workshop has referred to \textbf{AI for Software Engineering (AI4SE)} as an umbrella term to describe research that uses AI algorithms to tackle SE tasks \cite{SEI}.

Despite the unprecedented performance achieved by ML/DL models with higher complexity, they have been slow to be deployed in the SE industry. This reluctance arises due to prioritizing accuracy over \textbf{Explainability} -- AI systems are notoriously difficult to understand for humans because of their complex configurations and large model sizes \cite{DBLP:journals/csur/GuidottiMRTGP19}. From the perspective of the model user, explainability is needed to establish trust when imperfect \emph{``black-box''} models are used. For instance, a developer may seek to comprehend the rationale behind a DL-based vulnerability detection model's decision, i.e., why it predicts a particular code snippet as vulnerable, to facilitate analyzing and fixing the vulnerability \cite{SVulD,VulTeller}. For the model designers, explainability is required to investigate failure cases and direct the weak AI models in the proper paths as intended \cite{DBLP:conf/sigsoft/SunejaZZLM21}. In other words, merely a simple decision result (e.g., a binary classification label) without any explanation is often not good enough. This fact stimulates the urgent demand for designing algorithms capable of explaining the decision-making process of black-box AI models, leading to the creation of a novel research topic termed \textbf{eXplainable AI (XAI)} \cite{DBLP:journals/csur/DwivediDNSRPQWS23}.

\noindent\textbf{Our Work.}
To effectively chart the most promising path forward for research on the explainability for AI4SE, we conducted a \textbf{Systematic Literature Review (SLR)} to bridge this gap, providing valuable insights to the community. In this paper, we collected 108 primary studies published in 27 flagship conferences and journals over the last 13 years (2012-2024). The detailed process of our SLR can be found in the online appendix\footnote{\url{https://github.com/RISS-Vul/xai4se-paper/blob/master/Appendix.pdf}}. We focus on investigating the following \textbf{Research Questions (RQs)}:

\begin{itemize}[leftmargin=2em]
\item \textbf{\emph{RQ$_1$: What types of AI4SE studies have been explored for explainability?}}
\\
\textit{\underline{\textbf{Findings:}} (1) Primary studies can be categorized into 23 unique SE tasks across five major activities within software development life cycle; (2) Early studies predominantly concentrated on traditional classification tasks like Bug/Defect Prediction. Since 2021, researchers have turned their attention to a greater number of more complex SE tasks; (3) While there has been a recent wealth of work, there are still underrepresented topics in software requirements \& design and software management that should be considered by the SE community, suggesting a potential area of focus for future research in this field.}
\item \textbf{\emph{RQ$_2$: How XAI techniques are used to support SE tasks?}}
\begin{itemize}[leftmargin=1em]
\item[-] \emph{\textbf{RQ$_{2a}$:} What types of XAI techniques are employed to generate explanations?}
\item[-] \emph{\textbf{RQ$_{2b}$:} What format of explanation is provided for various SE tasks?}
\end{itemize}

\textit{\underline{\textbf{Findings:}} (1) Existing XAI techniques for SE tasks are mainly developed along five directions, including Out-of-the-Box Toolkit, Interpretable Model, Domain Knowledge, Attention Mechanism, as well as a set of other highly tailored approaches. The most popular of the five being Out-of-the-Box Toolkit, contributing $\approx$34\% of our surveyed studies; (2) Key factors guiding the selection encompass Task Fitness, Model Compatibility, and Stakeholder Preference; (3) A number of explanation formats have been explored in our surveyed studies, with the main formats utilized being Numeric, Text, Visualization, Source Code, and Rule. They were often tightly associated with a given SE task.}

\item \textbf{\emph{RQ$_3$: How well do XAI techniques perform in supporting various SE tasks?}}
\begin{itemize}[leftmargin=1em]
\item[-] \emph{\textbf{RQ$_{3a}$:} What baseline techniques are used to evaluate XAI4SE approaches?}
\item[-] \emph{\textbf{RQ$_{3b}$:} What benchmarks are used for these comparisons?}
\item[-] \emph{\textbf{RQ$_{3c}$:} What evaluation metrics are employed to measure XAI4SE approaches?}
\end{itemize}

\textit{\underline{\textbf{Findings:}} (1) In light of a notable scarcity of well-documented and reusable baselines or benchmarks, approximately 28.7\% of the benchmarks employed in the evaluations of our studied approaches were self-generated, with a significant portion not being publicly accessible or reusable; (2) There is no consensus on evaluation strategies for XAI4SE studies, and in many cases, the evaluation is only based on specific properties, such as correctness and coherence, or researchers' subjective intuition of what constitutes a good explanation.}
\end{itemize}

\noindent\textbf{Contributions.}
We anticipate that our findings will be instrumental in guiding future advancements in this rapidly evolving field. This study makes the following contributions:

%Although there have been a number of surveys on XAI methods \cite{XAI1,XAI2,XAI4}, these papers are targeted towards general XAI, and are not specific to the SE community. Hence, we conduct a review on explainability techniques exclusively related to SE tasks. Recently, the advancements in \textbf{Explainable Artificial Intelligence (XAI)} field has attracted researchers to develop explainability techniques for AI-based SE models.

\begin{itemize}[leftmargin=2em]
\item We present a systematic review of recent 108 primary studies on the topic of explainability for machine/deep learning-based software engineering, and pinpoint several potential directions for researchers and practitioners.
\item We describe the key applications of XAI4SE encompassing a diverse range of 23 unique SE tasks, grouped into five core SE activities within software development life cycle.
\item We synthesize a taxonomy of XAI techniques used in SE from an integration perspective, and analyze frequently used formats of explanation.
\item We summarize common evaluation means adopted by XAI4SE research, including available baselines, prevalent benchmarks, and commonly employed evaluation metrics, to determine their validity.
\item We discuss key challenges that using XAI techniques encounters within the SE field, and provide several practical guidelines for future research.
\item We maintain an interactive website, \url{https://riss-vul.github.io/xai4se-paper/}, with all of our data and results for reproducibility, and encourage contributions from the community to continue to push forward XAI4SE research.
\end{itemize}

\begin{table}[t]
 \caption{Comparison of Our Work with Previous Surveys/Reviews on Explainability for AI4SE Research}
  \centering
  \begin{threeparttable}
  \scalebox{0.69}{
  \begin{tabular}{|l|c|c|c|c|c|c|c|c|}
    \toprule
    \multirow{2}*{\textbf{Reference}} & \multirow{2}*{\textbf{Studied Model}} & \multirow{2}*{\textbf{Scope}} & \multirow{2}*{\textbf{\# Papers}} & \multirow{2}*{\textbf{Taxonomy}} & \multicolumn{3}{c|}{\textbf{Evaluation$^\dag$}} & \multirow{2}*{\textbf{Guideline}} \\
    ~ & ~ & ~ & ~ & ~ & \textbf{\rotatebox{0}{Baseline}} & \textbf{\rotatebox{0}{Benchmark}} & \textbf{\rotatebox{0}{Metric}} & ~\\
    \midrule
    Mohammadkhani et al. \cite{DBLP:journals/corr/abs-2302-06065} & ML \& DL & -2022 & 24 & General & \Circle  & \LEFTcircle & \LEFTcircle & \Circle \\
    Yang et al. \cite{yangzhou} & CodeLMs & 2019-2023 & 146 (16) & General & \Circle & \Circle & \Circle & \Circle \\
    \midrule
    \textbf{Our survey} & ML \& DL (+LLM) & 2012-2024 & 108 & Customized & \CIRCLE & \CIRCLE & \CIRCLE & \CIRCLE\\
    \bottomrule
  \end{tabular}}
  \begin{tablenotes}
        \footnotesize
        \item[$\dag$] $\Circle$ denotes not cover, $\LEFTcircle$ denotes partial cover, and $\CIRCLE$ denotes fully cover.
  \end{tablenotes}
  \end{threeparttable}
  \label{RelatedSurvey}
\end{table}

\noindent\textbf{Comparison with Existing Surveys.}
%With the growing need for transparency and accountability, both the XAI community and downstream fields, such as healthcare \cite{XAI4Medical2} and finance \cite{DBLP:journals/corr/abs-2309-11960}, have pushed forward substantial efforts in recent years to render black-box AI models more explainable. 
Recently, the SE community has embarked on a series of research activities regarding explainability, where several existing literature reviews or surveys \cite{DBLP:journals/corr/abs-2302-06065,yangzhou} have been produced, as summarized in Table \ref{RelatedSurvey}. Mohammadkhani et al. \cite{DBLP:journals/corr/abs-2302-06065} conducted a seminal survey on explainable AI for SE research. They primarily focus on the explainability of AI4SE models over the \emph{which}, the \emph{what}, and the \emph{how} dimensions, i.e., \emph{which} SE tasks are being explained, \emph{what} types of XAI techniques are adopted, and \emph{how} they are evaluated. Yang et al. \cite{yangzhou} reviewed 146 studies to investigate how non-functional properties of \textbf{Code Language Models (CodeLMs)}, including robustness, security, privacy, explainability, efficiency, and usability, were evaluated and enhanced. Despite the similarity in terms of the high-level topic, there remain some fundamental differences. First, they either focused narrowly on a single model architecture (e.g., CodeLMs), or only analyzed a small fraction of relevant literature published until June 2022. As a result, insights derived from them may not be generalizable to \emph{all} AI4SE solutions, or not keep pace with the ongoing development of the community. Second, they directly borrowed the general taxonomy, i.e., ante- and post-hoc explanation, from the XAI field to classify explanation techniques in SE tasks. This taxonomy is coarse-grained and may not be applicable to stakeholders with distinct objectives and expertise. Third, they did not (or only partially) explicitly discuss the evaluation aspects of reviewed papers, including available baselines, prevalent benchmarks, and commonly employed evaluation metrics. The lack of comprehensive evaluation may pose obstacles to readers interested in deploying XAI techniques in practical SE scenarios. Overall, by systematically reviewing publications from 2012 to 2024, spanning 13 years of research, we synthesize a detailed research roadmap of past work on XAI4SE, complete with identified open challenges and best guidelines for applying XAI techniques to SE tasks.

%Substantial efforts have been underway in recent years to enhance the explainability of black-box models. Simultaneously, the SE community also embarks on a series of research activities regarding explainability.

%These approaches operate either by analyzing the impact of specific regions of the input on the final prediction \cite{SHAP,LIME}, or by inspecting the network activation \cite{DBLP:journals/ijcv/SelvarajuCDVPB20,DBLP:conf/cvpr/ZhouKLOT16}. Table \ref{RelatedSurvey} summarizes some of their successful applications in downstream fields, such as healthcare \cite{XAI4Medical2} and finance \cite{DBLP:journals/corr/abs-2309-11960}. 

%This effort has led to advancements across a range of tasks including defect prediction \cite{Pyexplainer,DBLP:conf/icse/MahbubSR23}, vulnerability detection \cite{DBLP:conf/issta/HuWLPWZ023,DBLP:journals/tosem/ZouZXLJY21}, code summarization \cite{DBLP:journals/tse/JiangSWCYZ23,DBLP:conf/iwpc/GengWDWCZJ23}, malware detection \cite{DBLP:journals/tosem/WuCGFLWL21,DBLP:conf/issre/LiuT0022}, among many others \cite{WheCha,KGXQR,DBLP:conf/icse/CitoDMC22}. 

\noindent\textbf{Paper Organization.} The remainder of this paper is structured as follows: Section \ref{XAIPre} describes the preliminaries of XAI. The succeeding Sections \ref{RQ1}-\ref{RQ3} are devoted to answering each of these RQs individually. Section \ref{ChaAndOpp} discusses the challenges that still need to be solved and points out potential research opportunities. Section \ref{Guideline} outlines guidelines for conducting future work on XAI4SE based upon the findings of our SLR. Section \ref{Conclusion} concludes this paper.

\section{Explainable Artificial Intelligence: Preliminaries}\label{XAIPre}
This section first details several critical terminologies commonly used in the XAI field. Then, we offer a general overview of the taxonomy of XAI approaches, aiming to furnish the reader with a solid comprehension of this topic.

\subsection{Definition}
The greatest challenge in establishing the concept of XAI in SE is the ambiguous definition of \emph{interpretability} and \emph{explainability}. Those terms, together with \emph{interpretation} and \emph{explanation}, are often used interchangeably in the literature \cite{DBLP:journals/jair/BurkartH21,DBLP:conf/eurosp/NadeemVCPDBV23}. For example, quoting Doshi-Velez and Kim et al. \cite{Doshi-Velez}, interpretability is the ability ``to explain or to present in understandable terms to a human.'' By contrast, according to Lent et al. \cite{DBLP:conf/aaai/LentFM04}, an explainable AI means it can ``present the user with an easily understood chain of reasoning from the user’s order, through the AI’s knowledge and inference, to the resulting behavior.'' Some argue that the terms are closely related but distinguish between them, although there is no consensus on what the distinction exactly is \cite{XAI1,10.1145/3583558}. To ensure that we do not exclude work because of different terminologies, we equate them (and use them interchangeably) to keep a general, inclusive discussion regardless of this debate. In this survey, we frame explanations in the context of SE research using ML/DL and adopt the phrasing of Dam et al. \cite{DBLP:conf/icse/Dam0G18} as follows:
\begin{tcolorbox}[enhanced,breakable=true,skin=enhancedmiddle,borderline={1mm}{0mm}{MidnightBlue}]
	\textit{\textbf{Definition 1:} Explainability or Interpretability of an AI-powered SE model measures the degree to which a human observer can understand the reasons behind its decision (e.g. a prediction).}
\end{tcolorbox}

Under this context, there are two distinct ways of achieving explainability: (\ding{182}) making the entire decision-making process transparent and comprehensible (i.e., white-box/interpretable models); and (\ding{183}) explicitly providing an explanation for each decision (i.e., surrogate models). In addition, since SE is task-oriented, explanations in SE tasks should be viewed from a perspective that values practical use \cite{DBLP:journals/mima/Winter10}. As we observed in Section \ref{RQ2b}, there are multiple legitimate types of explanations for SE practitioners who have different intents and expertise.
% \begin{figure}
%   \centering
%   \includegraphics[width=.5\linewidth]{Figure/Tax.pdf}
% \caption{General taxonomy of the survey in terms of scope, stage, and portability.}
% \label{taxFig}
% \Description{}
% \end{figure}
\subsection{Taxonomy}\label{Taxonomies}
\begin{wrapfigure}{r}{.5\textwidth}
    \vspace{-5mm}
    \centering
    \includegraphics[width=.5\textwidth]{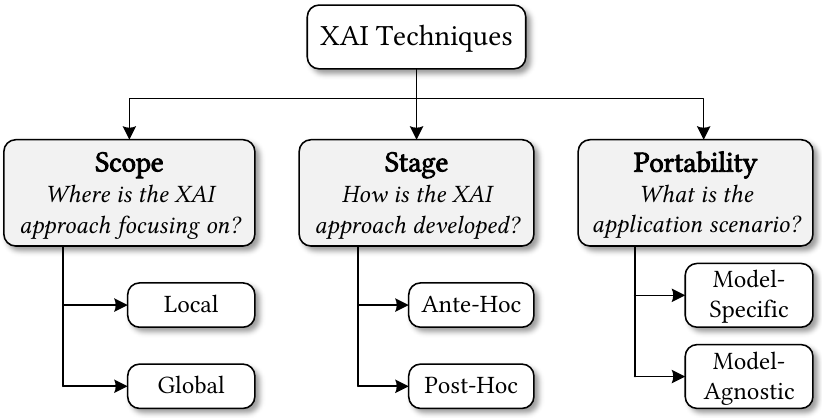}
    \caption{General taxonomy of the survey in terms of scope, stage, and portability.}\label{taxFig}
    \vspace{-2mm}
\end{wrapfigure}
Taxonomy is a useful tool to get an overview of the emerging field. Based on the previous literature \cite{speith2022review}, most XAI approaches can be categorized according to three criteria: (\ding{182}) \emph{scope} (local vs. global); (\ding{183}) \emph{stage} (ante-hoc vs. post-hoc); (\ding{184}) and \emph{portability} (model-specific vs. model-agnostic), as illustrated in Figure \ref{taxFig}.
%illustrates the general taxonomy of the XAI techniques, and each category is detailed in the following.

\noindent\textbf{Classification by Scope.}
The scope of explanations can be categorized as either \emph{local} or \emph{global} (some approaches can be extended to both) according to whether the explanations provide insights about the model functioning for the general data distribution or for a specific data sample, respectively. Local explainability approaches, such as LIME \cite{LIME} and SHAP \cite{SHAP}, seek to explain why a model performs a specific prediction for an individual input. Global explainability approaches work on an array of inputs to give insights into the overall behavior of the black-box model. Various rule-based models such as decision trees are in this category.

%Representative approaches include LIME \cite{LIME}, SHAP \cite{SHAP}, and Grad-CAM \cite{DBLP:journals/ijcv/SelvarajuCDVPB20}. Global explainability approaches work on an array of inputs to give insights into the overall behavior of the black-box model. 

%Various rule-based models such as decision trees are in this category.

%For example, given a file of interest and a defect prediction model, a locally explainable model might generate attribution scores for each code token in the file \cite{DBLP:journals/tse/Wattanakriengkrai22}. 
\noindent\textbf{Classification by Stage.}
XAI can be categorized based on whether the explanation mechanism is inherent within the model's internal architecture or is implemented following the model's learning/development phase. The former is named \emph{ante-hoc explainability} (also known as \emph{intrinsic explainability} or \emph{self-explainability}), while the latter refers to \emph{post-hoc explainability}. Most inherently interpretable approaches are model-specific such that any change in the architecture will need significant changes in the approach itself. By contrast, post-hoc approaches typically operate by perturbing parts of the data in a high-dimensional vector space to discern the contributions of various features to the model's predictions, or by analytically ascertaining the influence of different features on the prediction outcomes.

%By definition, 
%On the other hand, significant research interest in recent years is seen in developing post-hoc explanations as they can explain a well-trained black-box model decision without sacrificing the accuracy. 

\noindent\textbf{Classification by Portability.}
According to the models they can be applied to, explanation approaches can be further classified as \emph{model-specific} and \emph{model-agnostic}. Model-specific approaches require access to the internal model architecture, meaning that they are restricted to explain only one specific family of models. For example, Deconvolution \cite{DBLP:conf/eccv/ZeilerF14} is model-specific due to its ability to only explain the CNN model. Conversely, model-agnostic approaches can be used to explain arbitrary models without being constrained to any particular model architecture. 

Unlike other works that focus on underlying models and architectures, in this SLR, we design our taxonomy criteria with greater emphasis on the integration perspective of XAI and SE (Section \ref{RQ2}), making them better suited to the requirements of the SE community.

\section{RQ\texorpdfstring{$_1$}:: What types of AI4SE studies have been explored for explainability?}\label{RQ1}

\begin{table*}[t]
 \caption{Distribution of SE Tasks over Five SE Activities}
  \centering
  \scalebox{0.68}{
  \begin{tabular}{|c|c|l|}
    \toprule
    \textbf{SE Activity} & \textbf{SE Task (\# Papers)} & \textbf{References} \\
    \midrule
    \makecell[c]{Software Requirements \\ \& Design (Section \ref{Section:Requ})} & Requirement Classification (1) & \cite{DBLP:conf/re/DalpiazDAC19}\\
    \midrule
    \multirow{5}*{\shortstack{Software Development \\ (Section \ref{Section:Deve})}} & Code Understanding (11) & \cite{WheCha,DBLP:journals/tosem/YangFDWGW23,DBLP:conf/icse/WanZZSXJ22,DBLP:conf/sigsoft/RabinHA21,DBLP:journals/corr/abs-2308-12415,DBLP:conf/icse/CitoDMC22,DBLP:conf/kbse/LiZZWLXCJ23,DBLP:journals/tse/NaderPalacioVCRMP24,weima,DBLP:journals/jss/AwalR24,DBLP:journals/tse/PaltenghiPHZ24} \\
    ~ & Program Synthesis (7)  & \cite{DBLP:conf/pldi/EllisWNSMHCST21,DBLP:journals/pacmpl/NazariHSRR23,DBLP:conf/nips/Dang-Nhu20,DBLP:journals/tosem/LiuTLL24,DBLP:journals/pacmse/Chen00024,DBLP:conf/re/NorthAB24,DBLP:journals/pacmse/KouCW0024} \\
    ~ & Code Summarization (3)   &  \cite{DBLP:conf/iwpc/GengWDWCZJ23,DBLP:conf/kbse/PaltenghiP21,DBLP:journals/tse/JiangSWCYZ23}\\
    ~ & Code Search (2) & \cite{10.1145/3593800,DBLP:conf/kbse/WanSSXZ0Y19} \\
    ~ & API Recommendation (1) & \cite{DBLP:journals/tse/HuangLXZPXL24} \\
    \midrule
    \multirow{4}*{\shortstack{Software Testing \\ (Section \ref{Section:Test})}} &  Test Case-Related (5)& \cite{DBLP:conf/icse/YuLSR00L0W22,DBLP:conf/sigsoft/SunXTDLWZCN23,DBLP:conf/issre/AdigunHCF23,DBLP:conf/qrs/KeWFGS20,DBLP:journals/tosem/JodatCNS24}\\
    ~ & Debugging (4) & \cite{DBLP:conf/sigsoft/CitoD0M021,DBLP:conf/icse/GesiSGCA23,DBLP:conf/sigsoft/LampelJAZ21,DBLP:conf/sigsoft/KampmannHSZ20} \\
    ~ & Vulnerability Detection (14) & \cite{VulTeller,CoLeFunDa,SVulD,DBLP:conf/sigsoft/SunejaZZLM21,IVDETECT,DBLP:journals/tosem/ZouZXLJY21,DBLP:conf/icse/SunXLXZHZ23,VulExplainer,DBLP:conf/ijcai/LiuQWZHJ21,DBLP:conf/kbse/0007LY0CWM22,DBLP:conf/issta/HuWLPWZ023,DBLP:conf/icse/Cao0W0B0024,DBLP:conf/issta/Chu00W0S0024,DBLP:journals/tosem/ChengZWWBFGMW24} \\
    ~ & Bug/Fault Localization (3) & \cite{DBLP:conf/iwpc/WidyasariPHTZ022,DBLP:conf/wcre/WidyasariANS024,DBLP:journals/pacmse/KangAY24} \\
    \midrule
    \multirow{9}*{\shortstack{Software Maintenance \\ (Section \ref{Section:Main})}} & Program Repair (2)   & \cite{DBLP:conf/msr/MarkovtsevLMSB19,DBLP:conf/qrs/BoH0JW24} \\
    ~ & Malware/Anomaly Detection (12)  & \cite{DBLP:journals/tosem/WuCGFLWL21,DBLP:conf/wcre/LiCZSP22,DBLP:conf/issre/LiuT0022,DBLP:conf/sigsoft/ZhaoWLPWPWFWZSP21,DBLP:conf/icse/ZengZXMQZCZZZGFRLZ23,DBLP:conf/qrs/WangYH21,webTamper,DBLP:conf/iwpc/0001TM0MCZYJ24,DBLP:conf/icse/HrustoRO24,DBLP:conf/sigsoft/ZhaoCWPWWZFNZSP20,DBLP:journals/tosem/LyuRLCJ22,DBLP:journals/tosem/LiJLHHHZWC20} \\
    ~ & Bug/Defect Prediction (19)  & \cite{DBLP:conf/icse/MahbubSR23,Pyexplainer,DBLP:journals/ase/SantosFVVZ20,DBLP:journals/tse/Wattanakriengkrai22,DBLP:journals/tse/RajapakshaTJBGB22,DBLP:conf/wcre/YangZZZX23,DBLP:journals/jss/ZhengSCD22,DBLP:journals/ese/MoriU19,DBLP:journals/tse/JiarpakdeeTDG22,DBLP:conf/wcre/LeeL23,DBLP:conf/msr/GaoZY22,DBLP:journals/infsof/GaoZZ22,FOX,DBLP:conf/apsec/ShinANWHW23,DBLP:journals/jss/YangZZXZQ24,DBLP:conf/sigsoft/ChenFKM18,DBLP:journals/tse/LinTH22,DBLP:journals/tse/PengM22,DBLP:conf/msr/JiarpakdeeTG21} \\
    ~ & OSS Sustainability Prediction (1) & \cite{DBLP:conf/sigsoft/Xiao0XZ023}\\
    ~ & Root Cause Analysis (5) & \cite{DBLP:conf/sigsoft/DingZWXMWZCGGFR23,DBLP:conf/sigsoft/YaoPCWSJXNP24,deephunt,slim,DBLP:conf/sigsoft/LiZ0LWCNCZSWDDP22}\\
    ~ & Code Review (2) & \cite{DBLP:conf/sigsoft/YangXZZB23,DBLP:conf/esem/SarkerSWB23} \\
    ~ & Code Smell Detection (6) & \cite{DBLP:conf/kbse/WangLLCLW20,DBLP:journals/tosem/RenXXLWG19,DBLP:conf/compsac/YinSZ21,DBLP:journals/tse/TsoukalasMAACK24,DBLP:journals/jss/HuangYFSZL24,TODO} \\
    ~ & Code Clone Detection (1) & \cite{DBLP:conf/apsec/AbidCJ23} \\
    ~ & Bug Report-Related (5)  & \cite{DBLP:journals/ese/ShihabIKIOAHM13,DBLP:conf/qrs/DingFYH21,DBLP:conf/issre/HeXF0YL20,DBLP:journals/ese/LaiqABE24,DBLP:journals/ese/SchulteLH24} \\
    \midrule
    \multirow{4}*{\shortstack{Software Management \\ (Section \ref{Section:Mana})}} &  Mining Software Repositories (1) & \cite{KGXQR} \\
    ~ & Configuration Extrapolation (1) & \cite{DBLP:conf/sigsoft/0006PCH21} \\
    ~ & Effort/Cost Estimation (1)  & \cite{GPT2SP} \\
    ~ & Developer Recommendation (1)  & \cite{DBLP:journals/tse/XieYWH22} \\
    \bottomrule
  \end{tabular}}
  \label{dataset}
\end{table*}

This RQ aims to investigate the application scenarios of XAI techniques in helping improve the explainability of various AI4SE models. In total, we identified 23 separate SE tasks where an XAI technique had been applied. These tasks span across five main phases of \textbf{Software Development Life Cycle (SDLC)} \cite{DBLP:journals/sigsoft/Ruparelia10a}. The full taxonomy is displayed in Table \ref{dataset}, which associates the relevant primary study paired with the SE task \& activity it belongs to.

%Out of the \textcolor{red}{108} primary studies we analyzed for this SLR, 

%, including \textcolor{red}{software requirements \& design}, software development, software testing, software maintenance, and software management

\subsection{How XAI Are Used in Specific SE Tasks?}
In this subsection, we delved into the progress of various SE tasks that applied XAI techniques. By investigating this RQ, we aimed to obtain a clear understanding of what has been done and what else can be done.

%in advancing practices for explainable AI4SE solutions.

\subsubsection{SE Tasks in Software Requirements \& Design}\label{Section:Requ}

Software requirements refer to specific descriptions of conditions or capabilities needed by users, systems, or system components, while software design involves the process of defining the structure, components, functionalities, interfaces, and their relationships within a software system. During this phase, only one topic, i.e., \emph{Requirement Classification}, is explored, leaving ample space for further exploration.

\noindent\textbf{Requirement Classification.}
As a key example of ML applied to requirements engineering, requirement classification aims to categorize software requirements into different classes or types, such as functional and non-functional requirements. Dalpiaz et al. \cite{DBLP:conf/re/DalpiazDAC19} constructed ML classifiers based on more general linguistic features (e.g., dependency types), and leveraged modern rule-based XAI tools to identify those features that appeared commonly and that helped distinguish functional and quality aspects.

\subsubsection{SE Tasks in Software Development}\label{Section:Deve}
There are wide-ranging applications of XAI techniques in software development, encompassing tasks such as \emph{Code Understanding}, \emph{Program Synthesis}, and \emph{Code Summarization}.

\noindent\textbf{Code Understanding.}
Code understanding refers to the process of comprehending and analyzing source code deeply. Within the context of data-driven SE research, code understanding aims to seek an effective way to map source code into high-dimensional semantic space, thereby supporting a variety of code-centric downstream tasks. Inspired by the capability of complex AI models, deep neural networks in particular, in learning rich representations of raw data, a series of code models are trained on labeled (e.g., CodeSearchNet \cite{husain2019codesearchnet}) or unlabeled code corpus (e.g., CodeXGlue \cite{DBLP:conf/nips/LuGRHSBCDJTLZSZ21}). This training process produces code embeddings with rich contexts and semantics. Yang et al. \cite{DBLP:journals/tosem/YangFDWGW23} proposed Graph Tensor Convolution Neural Network (GTCN), a novel code representation learning model which is capable of comprehensively capturing the distance information of code sequences and structural code semantics, to generate accurate code embeddings. GTCN was self-explainable because the tensor-based model reduced model complexity, which was beneficial for capturing the data features from the simpler model space. Wan et al. \cite{DBLP:conf/icse/WanZZSXJ22} proposed three types of structural analysis, including attention analysis, probing on the word embedding, and syntax tree induction, to explore why the pre-trained language models work and what they indeed capture in SE tasks.

\noindent\textbf{Program Synthesis.}
Program synthesis refers to the automated process of generating source code or software programs to meet specified requirements based on specified specifications or requirements. Although DL-based program synthesis applications, such as CodeGeeX\footnote{\url{https://codegeex.cn/en-US}} and GitHub Copilot\footnote{\url{https://copilot.microsoft.com/}}, have integrated into the daily workflow of software developers around the globe due to their high, even human-competitive accuracy, lacking transparency makes the automatically generated programs untrustworthy. To explain the fundamental working mechanism, Chen et al. \cite{DBLP:journals/pacmse/Chen00024} proposed a causality-inspired approach, which constructed a dependency graph to capture not only the natural dependency between input and output tokens, but the dependency among output tokens as well. Then, they leveraged causal inference to quantify the contribution of each dependency edge to the end prediction result. Besides, Nazari et al. \cite{DBLP:journals/pacmpl/NazariHSRR23} introduced subspecifications as a general mechanism to identify the constraints imposed by the global specification on individual parts of the implementation as explanatory notes.

\noindent\textbf{Code Summarization.}
Code summarization, also known as \emph{code comment generation}, is a \emph{code-to-text} task that attempts to automatically generate textual descriptions directly from source code. A promising solution is \textbf{Neural Machine Translation (NMT)} \cite{NMT1,NMT2}. Despite their effectiveness, such end-to-end summarization without any explanation is still far from being usable in practice. To this end, Geng et al. \cite{DBLP:conf/iwpc/GengWDWCZJ23} designed two types of explanation strategies applicable to different application scenarios to identify the corresponding code parts used to generate summarization. The black-box explanation strategy aimed to localize the code segment that sensitive to program mutations, while the white-box explanation strategy focused on inspecting the attention score of the each code token. Jiang et al. \cite{DBLP:journals/tse/JiangSWCYZ23} proposed CCLink, a model-independent framework which aimed to find the code segments that contribute to the generation of key information in the auto-generated comments. CCLink generated a series of code mutants from key phrases in the auto-generated comment, and tailored data mining algorithms to construct the links between code and its auto-generated comment. This in turn allowed CCLink to visualize links as the comment explanations to developers.

\noindent\textbf{Others.}
Apart from the above tasks, a number of studies applied XAI techniques on other research topics for improving AI-assisted software development. For instance, code search aims to retrieve source code that meets users' natural language queries from a large codebase. Wang et al. \cite{10.1145/3593800} proposed an explainable code search tool, namely XCoS, to bridge the knowledge gap between the query and candidate code snippets. Based on the background knowledge graph extracted from Wikidata and Wikipedia, XCoS provided conceptual association paths, relevant descriptions, and additional suggestions, as explanations. Furthermore, it designed an interactive User Interface (UI) which organized explanatory information in the form of trees to help developers intuitively understand the rationale behind the search results. Huang et al. \cite{DBLP:journals/tse/HuangLXZPXL24} designed a novel knowledge-aware HumanAI dialogue agent to interact with developers and return APIs with relevance explanation and extended knowledge.

\subsubsection{SE Tasks in Software Testing}\label{Section:Test}
Within the context of software testing, we found versatile applications of XAI techniques across a spectrum of tasks, including \emph{Test Case-Related Automation}, \emph{Debugging}, \emph{Vulnerability Detection}, and \emph{Bug/Fault localization}.

\noindent\textbf{Test Case-Related.}
The design and implementation of test cases occupy most of the software testing cycle because their quality directly affects the quality of software testing. In recent years, the combination of intelligent technology and test scenarios, such as test case generation \cite{DBLP:conf/icse/YuLSR00L0W22}, test case recommendation \cite{DBLP:conf/qrs/KeWFGS20}, and test case diversity analysis \cite{DBLP:conf/issre/AdigunHCF23}, has received widespread attention. Yu et al. \cite{DBLP:conf/icse/YuLSR00L0W22} improved the DL-based automated assertion generation approach by integrating the \textbf{Information Retrieval (IR)}-based assertion retrieval technique and the retrieved-assertion adaptation technique. The assertion retrieval using IR also yields the corresponding focal-test. In this context, focal-test refers to a pair consisting of a test method without an assertion and its focal method. Developers can then use this focal-test as a valuable reference during assertion inspection. To make full use of historical test cases to improve test efficiency, Ke et al. \cite{DBLP:conf/qrs/KeWFGS20} proposed an explainable test case recommendation approach based on the knowledge graph. Once a historical test case is predicted to be prone to revealing defects by the classifier, the corresponding knowledge chain in the software test knowledge graph will be returned as auxiliary information to help testers understand the reason for the recommendation.

\noindent\textbf{Debugging.}
ML/DL-based models for SE tasks, similar to traditional software, suffer from errors that result in unexpected behavior or incorrect functionality. Due to the black-box nature of complex AI models, the debugging process designed for traditional software, which involves reviewing the code, tracing abnormal execution flows, and isolating the root cause of the problem, may not be applicable to ML/DL-based SE models. Motivated by the idea that diagnostic features could be potentially useful, Cito et al. \cite{DBLP:conf/sigsoft/CitoD0M021} presented a model-agnostic rule induction technique to explain when a code model performed poorly. The misprediction diagnosis model aimed to learn a set of rules that collectively cover a large portion of the model’s mispredictions, each of which correlates strongly with model mispredictions. The evaluation results showed that these learned rules are both accurate and simple.

\noindent\textbf{Vulnerability Detection.}
Software vulnerabilities, sometimes called security bugs, are weaknesses in an information system, security procedures, internal controls, or implementations that could be exploited by a threat actor for a variety of malicious ends. As such weaknesses are unavoidable during the design and implementation of the software, and detecting vulnerabilities in the early stages of the software life cycle is critically important. Benefiting from the great success of DL in code-centric software engineering tasks, an increasing number of learning-based vulnerability detection approaches have been proposed. To reveal the decision logic behind the binary detection results (vulnerable or not), most efforts focus on searching for important code tokens that positively contribute to the model’s prediction. For example, Li et al. \cite{IVDETECT} leveraged GNNExplainer \cite{DBLP:conf/nips/YingBYZL19} to simplify the target instance to a minimal PDG sub-graph consisting of a set of crucial statements along with program dependencies while retaining the initial model prediction. Additionally, several approaches turn to providing explanatory descriptions to help security analysts understand the key aspects of vulnerabilities, including vulnerability types \cite{VulExplainer}, root cause \cite{DBLP:conf/icse/SunXLXZHZ23}, similar vulnerability reports \cite{SVulD}, and so on. Zhou et al. \cite{CoLeFunDa} proposed a novel contrastive learning framework based on a combination of unsupervised and supervised data augmentation strategy to train a function change encoder, and further fine-tuned three downstream tasks to identify not only silent vulnerability fixes, but also corresponding vulnerability types and exploitability rating.

\noindent\textbf{Bug/Fault Localization.}
Bug localization refers to the process of pinpointing the exact location in the codebase where the bug originates based on bug reports or issue descriptions provided by users or testers. From the perspective of enhancing the effectiveness of bug localization, Widyasari et al. \cite{DBLP:conf/iwpc/WidyasariPHTZ022} formulated the localization task as a binary classification problem, i.e., predicting whether a test case will fail or pass. They applied TreeSHAP \cite{TreeSHAP}, a local model-agnostic explanation technique, to identify which parts of code are important in each failed test case. From the perspective of providing explanations for localized bugs, Li et al. \cite{DBLP:conf/sigsoft/LiZ0LWCNCZSWDDP22} respectively designed the global and local explanation strategies to explain the model predictions. The global explanation strategy leveraged decision trees as the surrogate models to infer the decision paths leading to only faulty or normal failure units, while the local explanation strategy compared the incoming failure with each historical failure to explain how the model diagnosed a given failure.

\subsubsection{SE Tasks in Software Maintenance}\label{Section:Main}
Software maintenance is the process of changing, modifying, and updating software to keep up with customer needs. The applications of XAI in software maintenance are diverse, including \emph{Malware/Anomaly Detection}, \emph{Bug/Defect Prediction}, \emph{Root Cause Analysis}, \emph{Code Smell Detection}, \emph{Bug Report-Related Automation}, etc.

\noindent\textbf{Malware/Anomaly Detection.}
As the emerging collaborative software development modes of open source have become increasingly popular, the overall security risk trend, such as malicious applications (malware) and commits, in complex and intertwined software supply relationships increases significantly. Wu et al. \cite{DBLP:journals/tosem/WuCGFLWL21} proposed an explainable ML-based approach, named XMal, to not only predict whether an app is malware, but also unveil the rationale behind its prediction. For this purpose, XMal built a semantic database based on malware key features and functional descriptions in Android developer documentation\footnote{\url{https://developer.android.google.cn/docs}}, and leveraged the mapping relation between the malicious behaviors and their corresponding semantics to generate descriptions for malware.

\noindent\textbf{Bug/Defect Prediction.}
In the past few years, defect prediction is the most extensive and active research topic in software maintenance. According to different granularities, these studies can be further classified into two categories: file-level and commit-level (also known as \textbf{Just-In-Time (JIT)}) defect prediction. File-level defect prediction techniques often employ a set of hand-crafted feature metrics extracted from a software product to construct the classification model. For instance, Yang et al. \cite{DBLP:conf/wcre/YangZZZX23} proposed a weighted association rule based on the contribution degree of features to optimize the process of rule generation, ranking, pruning, and prediction. In addition, some studies directly adopt source code (e.g., code tokens) as meaningful semantic units for defect prediction. Wattanakriengkrai et al. \cite{DBLP:journals/tse/Wattanakriengkrai22} formalized defect explanation as a line-level prediction task, and used the model-agnostic technique LIME \cite{LIME} to identify risky code tokens in predicted defective files.

By contrast, JIT defect prediction task aims to help developers prioritize their limited energy and resources on the most risky commits that are most likely to introduce defects. Zheng et al. \cite{DBLP:journals/jss/ZhengSCD22} trained a random forest classifier based on six open-sourced projects as a JIT defect prediction model, and adopted LIME to identify crucial features. The evaluation experiments showed that the classifier trained on the five most important features of each project could achieve 96\% of the original prediction accuracy. Similarly, Pornprasit et al. \cite{Pyexplainer} randomly generated synthetic neighbors around the to-be-explained instance based on the actual characteristics of defect-introducing commits, and built a local rule-based regression model to explain the interactions between defect-introducing features.

\noindent\textbf{Root Cause Analysis.}
Despite the broad adoption of microservice architecture in modern production systems, it is inherently susceptible to critical production incidents, underscoring the importance of swiftly identifying and addressing the root causes of any issues to ensure quick system recovery. Prior studies directly fed multi-modal data to a deep neural network, suffering from limited interpretability and operability. Yao et al. \cite{DBLP:conf/sigsoft/YaoPCWSJXNP24} analyzed causal relationships of events transformed from multi-modal observation data, and provided interpretable parameters with clear physical meanings that align well with the operational experience of site reliability engineers, thereby facilitating the integration of their expertise directly into the analysis process through human feedback. Ding et al. \cite{DBLP:conf/sigsoft/DingZWXMWZCGGFR23} leveraged reinforcement learning to learn an interpretable pruning policy for the service dependency graph to automatically eliminates redundant components. The policy is based on graph pruning rules derived from experienced engineers and comprehensive trace analysis, ensuring domain knowledge and industry best practices are incorporated.

\noindent\textbf{Code Smell Detection.}
Due to the delivery date pressure or developer oversight, code smells may be introduced in software development and evolution, thereby reducing the understandability and maintainability of source code. One of the common code smells is \textbf{Technical Debt (TD)}, which reflects the trade-off software engineers make between short-term benefits and long-term stability. TD is accumulated when developers make wrong or unhelpful technical decisions, either intentionally or unintentionally, during software development. Despite the effectiveness of DL-based code smell detection approaches in automatically building the complex mapping between code features and predictions, they cannot provide the rationale for the prediction results. To understand the basis of the CNN-based detection model's decisions, Ren et al. \cite{DBLP:journals/tosem/RenXXLWG19} proposed a backtracking-based approach, which exploited the computational structure of CNN to derive key phrases and patterns from input comments.

\noindent\textbf{Bug Report-Related.}
A Bug report is one of the important crucial SE documentation for software maintenance. High-quality bug reports can effectively reduce the cost of fixing buggy programs \cite{DBLP:conf/kbse/BoJ000024}. For example, instead of using other black-box models, Shihab et al. \cite{DBLP:journals/ese/ShihabIKIOAHM13} employed an explainable ML model (i.e., decision tree) to understand which attributes affected whether or not a bug would be re-opened. To analyze the impact of test-related factors to DL-based bug report prediction models, Ding et al. \cite{DBLP:conf/qrs/DingFYH21} applied SHAP to compute the importance of tell smell features.

\noindent\textbf{Others.}
Some primary studies explored the explainability of certain special SE tasks. For example, Markovtsev et al. \cite{DBLP:conf/msr/MarkovtsevLMSB19} proposed a decision tree-based explainable model to express the found format patterns with compact human-readable rules. Xiao et al. \cite{DBLP:conf/sigsoft/Xiao0XZ023} leveraged the local explanations generated by LIME from the XGBoost model to analyze the contribution of variables to sustained activity in different project contexts.

\subsubsection{SE Tasks in Software Management}\label{Section:Mana}
There are four literature involving the utilization of XAI in software management, involving the following main SE tasks, i.e., \emph{Mining Software Repositories.}, \emph{Configuration Extrapolation}, \emph{Effort/Cost Estimation}, and \emph{Developer Recommendation}.

\noindent\textbf{Mining Software Repositories.}
As one of the most popular and widely-used technical Q\&A sites in SE communities, \textbf{Stack Overflow\footnote{\url{https://stackoverflow.com/}} (SO)} plays an increasingly important role in software development. When facing software programming problems such as implementing specific functionalities or handling errors in code, developers often turn to SO for help. To provide both accurate and explainable retrieval results, Liu et al. \cite{KGXQR} proposed KGXQR, a knowledge graph-based question retrieval approach for programming tasks. KGXQR constructed a software development-related concept knowledge graph to find the desired questions, and further generated explanations based on the association paths between the concepts involved in the query and the SO questions to bridge the knowledge gap.

\noindent\textbf{Configuration Extrapolation.}
Configuration management is a process in systems engineering for establishing consistency of a product’s attributes throughout its life. The increasing configurability of modern software also puts a burden on users to tune these configurations for their target hardware and workloads. To configure software applications efficiently, ML models have been applied to model the complex relationships between configuration parameters and performance. To understand the underlying factors that caused the low performance, Ding et al. \cite{DBLP:conf/sigsoft/0006PCH21} used an inherently interpretable linear regression model \cite{faraway2004linear} to find the configuration with the best predicted performance. They provided interactive visualization charts (e.g., radar charts, bar charts) to explain the relationships between the application-level configuration parameters and ultimate performance.

\noindent\textbf{Effort/Cost Estimation.}
Effort/Cost estimation is the process of predicting how much effort is required to complete a particular task or project. It is a crucial aspect of project management, playing a significant role in setting realistic timelines and allocating resources efficiently. A representative effort estimation activity is story point estimation, which is a regression task to measure the overall effort required to fully implement a product backlog item. Fu et al. \cite{GPT2SP} presented GPT2SP, a Transformer-based approach that captures the relationship among words while considering the context surrounding a given word and its position in the sequence. It is designed to be transferable to other projects while remaining explainable. They leveraged two concepts (i.e., feature-based explanations and example-based explanations) of XAI to 1) help practitioners better understand what are the most important word that contributed to the story point estimation of the given issue; and 2) search for the best supporting examples that had the same word and story point from the same project.

\noindent\textbf{Developer Recommendation.}
Collaboration efficiency is of paramount importance for software development. Although a lot of efforts in recommending suitable developers have been made in both research and practice in recent years, such approaches often suffer from low performance due to the difficulty of learning the developer’s expertise, willingness, relevance as well as the sparsity of explicit developer-task interactions. Xie et al. \cite{DBLP:journals/tse/XieYWH22} proposed a multi-relationship embedded approach named DevRec, in which they explicitly encoded the collaboration relationship, interaction relationship, and similarity relationship into the representation learnings of developers and tasks. DevRec also visualized the high-order connectivity and attentive embedding propagation in the recommendation subgraphs to explain why a task was recommended (or assigned) to the developer.

\subsection{Exploratory Data Analysis}
\begin{figure}[t]
        \centering
        \subcaptionbox{Distribution of XAI4SE studies in different SE activities.\label{RQ1A}}{
        \centering
        \includegraphics[width = .4\linewidth]{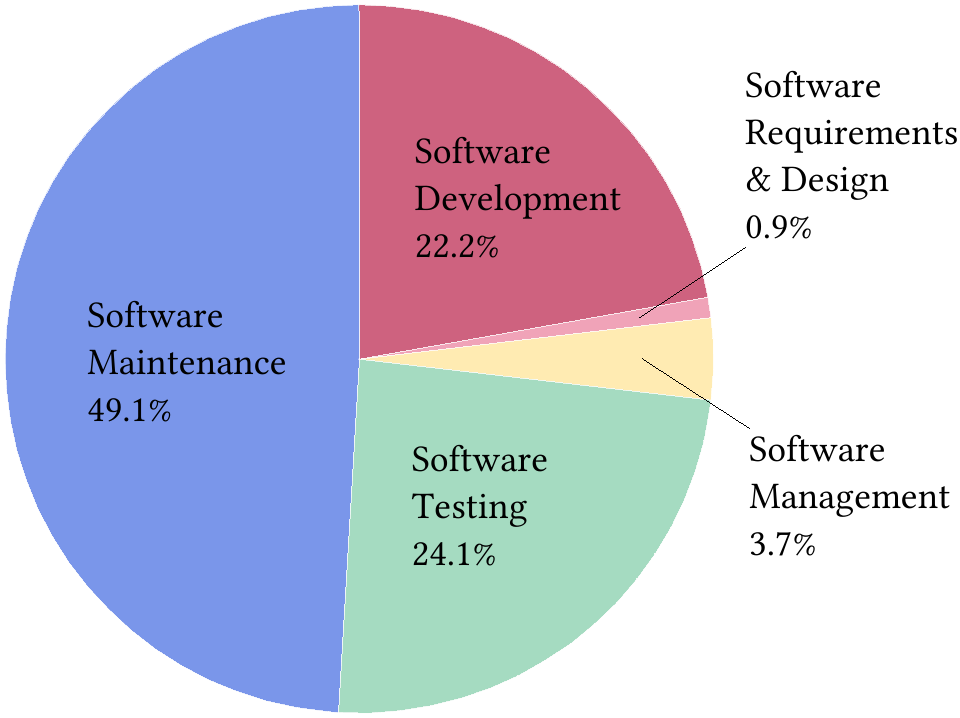}
    }
    \hspace{15mm}
        \subcaptionbox{Distribution of main contributions in different XAI4SE studies.\label{RQ1B}}{
        \includegraphics[width = .385\linewidth]{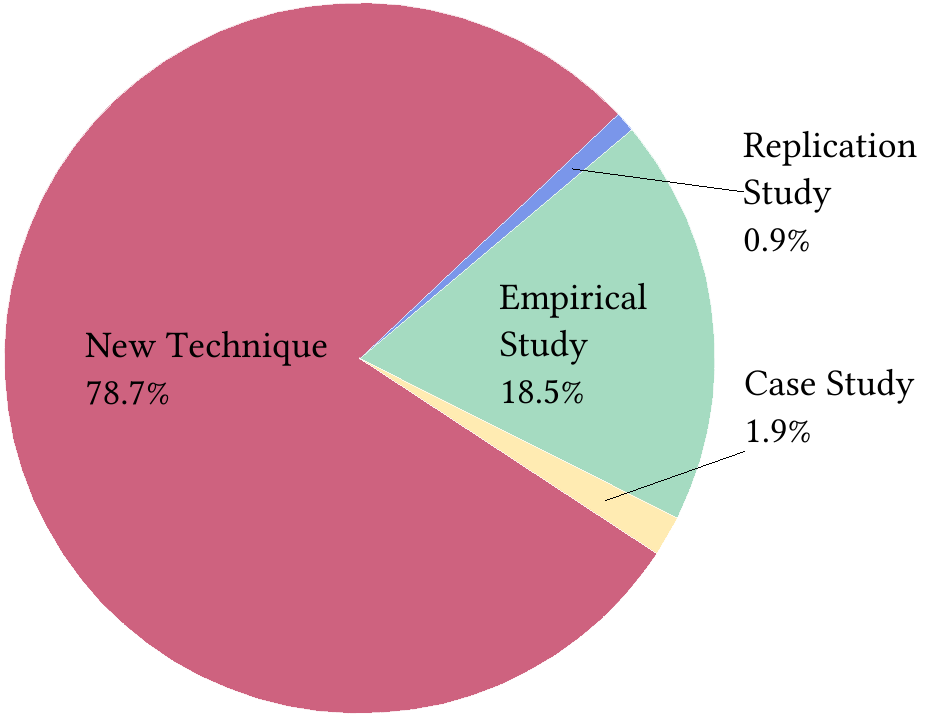}
    }
        \caption{Distribution of XAI4SE studies across different SE activities and contribution types.}
    \Description{}
    \end{figure}

Figure \ref{RQ1A} describes the distribution of 108 primary studies in five SE activities. It is noteworthy that the highest number of studies is observed in software maintenance, comprising 49.1\% of the total research volume. Following that, 24.1\% of studies were dedicated to software testing, and 22.2\% of studies focused on solving SE tasks in software development. This distribution underscores the vital focus on development and maintenance tasks. By contrast, software requirements \& design (0.9\%) and software management (3.7\%) only account for a marginal proportion of the research share, suggesting a relatively limited exploration in these areas. 
To further identify the main contribution of each primary study, we also investigated the contribution statements in each paper, and then grouped them into four categories, i.e., \emph{New Technique}, \emph{Empirical Study}, \emph{Case Study}, and \emph{Replication Study}. As shown in Figure \ref{RQ1B}, 78.7\% of the primary studies concentrated on proposing novel explanation techniques, while 18.5\% of the research focused on leveraging off-the-shelf XAI tools to empirically study the explainability of certain AI4SE solutions from different perspectives (e.g., stability \cite{DBLP:conf/apsec/ShinANWHW23}, consistency \cite{DBLP:journals/tosem/LyuRLCJ22}). Another two works \cite{DBLP:journals/corr/abs-2308-12415,DBLP:journals/ese/LaiqABE24} performed case studies (1.9\%) in real world, especially for enterprise usage, to investigate practitioners' adoption of AI4SE solutions. The remaining one \cite{DBLP:journals/jss/HuangYFSZL24} conducted a replication study (0.9\%) to validate the controversial conclusions in terms of local and global explanations.

\begin{figure}[t]
  \centering
  \includegraphics[width=.9\linewidth]{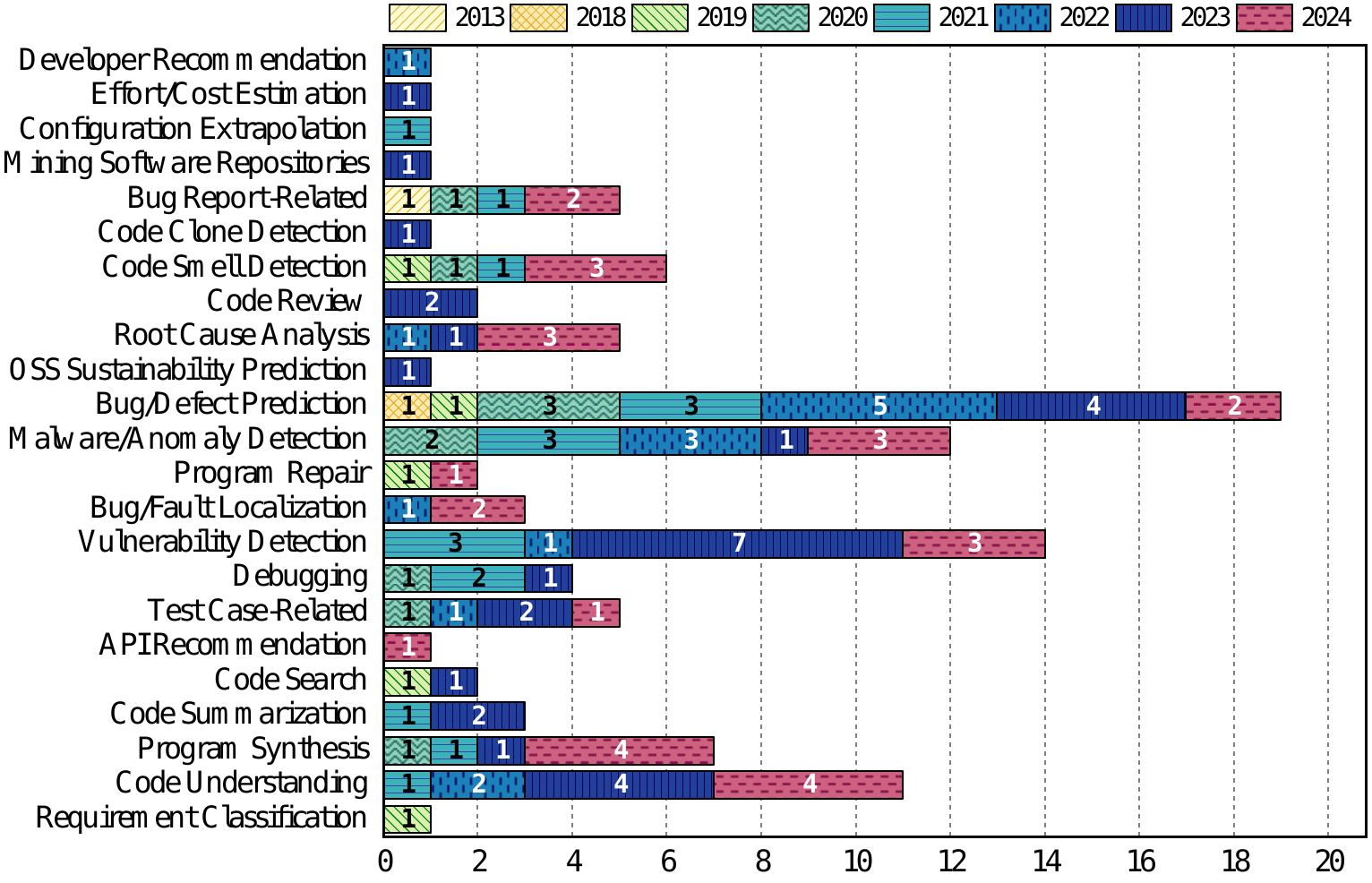}
\caption{Papers published per year according to SE tasks.}
\label{SETask}
\Description{}
\end{figure}

Figure \ref{SETask} displays a visual breakdown of these SE tasks. Unsurprisingly, there was very little work done between before the years of 2012 and 2018. Early SE tasks to explore the explainability were those of \emph{Bug Report-Related Automation} and \emph{Bug/Defect Prediction}. It was not until 2020 that the diversity experienced a significant increase, including tasks such as \emph{Malware/Anomaly Detection}, \emph{Debugging}, and \emph{Program Synthesis}. It is noteworthy that there are three main SE tasks that have consistently maintained high activity levels over the years: \emph{Bug/Defect Prediction}, \emph{Vulnerability Detection}, and \emph{Malware/Anomaly Detection}, composing $\approx$42\% of the studies we collected. We suspect that a variety of reasons contribute to the multiple applications of XAI in these tasks. First and foremost, is that these tasks are essentially binary classification problems that ML/DL models have shown promising results in. The second-largest reason is the mandatory requirement of high-stakes applications. In fact, black-box models are not even allowed in regulated fields unless they are supplemented with explanations \cite{DBLP:journals/aim/GoodmanF17}. In addition, several tasks (e.g., \emph{Mining Software Repositories} \cite{KGXQR}, \emph{Root Cause Analysis} \cite{deephunt,slim}, and \emph{Bug/Fault Localization} \cite{DBLP:conf/wcre/WidyasariANS024,DBLP:journals/pacmse/KangAY24}) that had yet to be explored or were underrepresented before \cite{DBLP:journals/corr/abs-2302-06065} have recently gained traction of the research community, in part due to their relative success in practice. We anticipate that this trend continues as more powerful AI4SE solutions evolve from experimental prototypes to practical tools.

%as prior work \cite{Memento,DBLP:conf/msr/JiarpakdeeTG21} do.} 
%Overall, We believe such an extension could lead to more practical and actionable AI-driven solutions across the entire SDLC.

%(e.g., dependability and security-related tasks as shown in ICSE'25's topics of interest\footnote{\url{https://conf.researchr.org/track/icse-2025/icse-2025-research-track}}) 

\begin{myhbox}{\ding{45} $\blacktriangleright$ RQ$_1$ - Summary $\blacktriangleleft$}
\begin{itemize}[leftmargin=2em]
\item We categorized a total of 108 primary studies into 23 unique SE tasks across five major activities within SDLC. Subsequently, we delved into the progress of existing XAI4SE research among these SE activities.
\item Attention of academics and practitioners has experienced a notable shift across a 13-year period. Early studies predominantly concentrated on traditional classification tasks like \emph{Bug/Defect Prediction}. Since 2021, the set of target SE topics grew to become more diverse and complex, including tasks such as \emph{Developer Recommendation}, \emph{Root Cause Analysis}, and \emph{Program Synthesis}.
\item While there has been a recent wealth of work, there are still underrepresented topics in software requirements \& design and software management that should be considered by the SE community, suggesting a potential area of focus for future research in this field.
\end{itemize}
\end{myhbox}

\section{RQ\texorpdfstring{$_2$}:: How XAI techniques are used to support SE tasks?}
In Section \ref{RQ1}, we analyzed which AI-assisted SE tasks have been explored for explainability to date. In this part, we turn our attention to two key components of XAI: \emph{explanation approaches} and \emph{explanation formats}. Establishing the association between explanation approaches and target SE tasks helps to empirically determine whether certain XAI techniques are particularly suitable for specific SE tasks. Meanwhile, the explanation formats adopted across different SE tasks reveal key aspects that the stakeholders seek to understand from the decision of a given black-box model. Specifically, we aimed to create a taxonomy of XAI techniques for AI4SE studies and determine if there was a correlation between the explanation approaches and explanation formats.

\subsection{RQ\texorpdfstring{$_{2a}$}:: What Types of XAI Techniques Are Employed to Generate Explanations?}\label{RQ2}
\begin{figure}
  \centering
  \includegraphics[width=.9\linewidth]{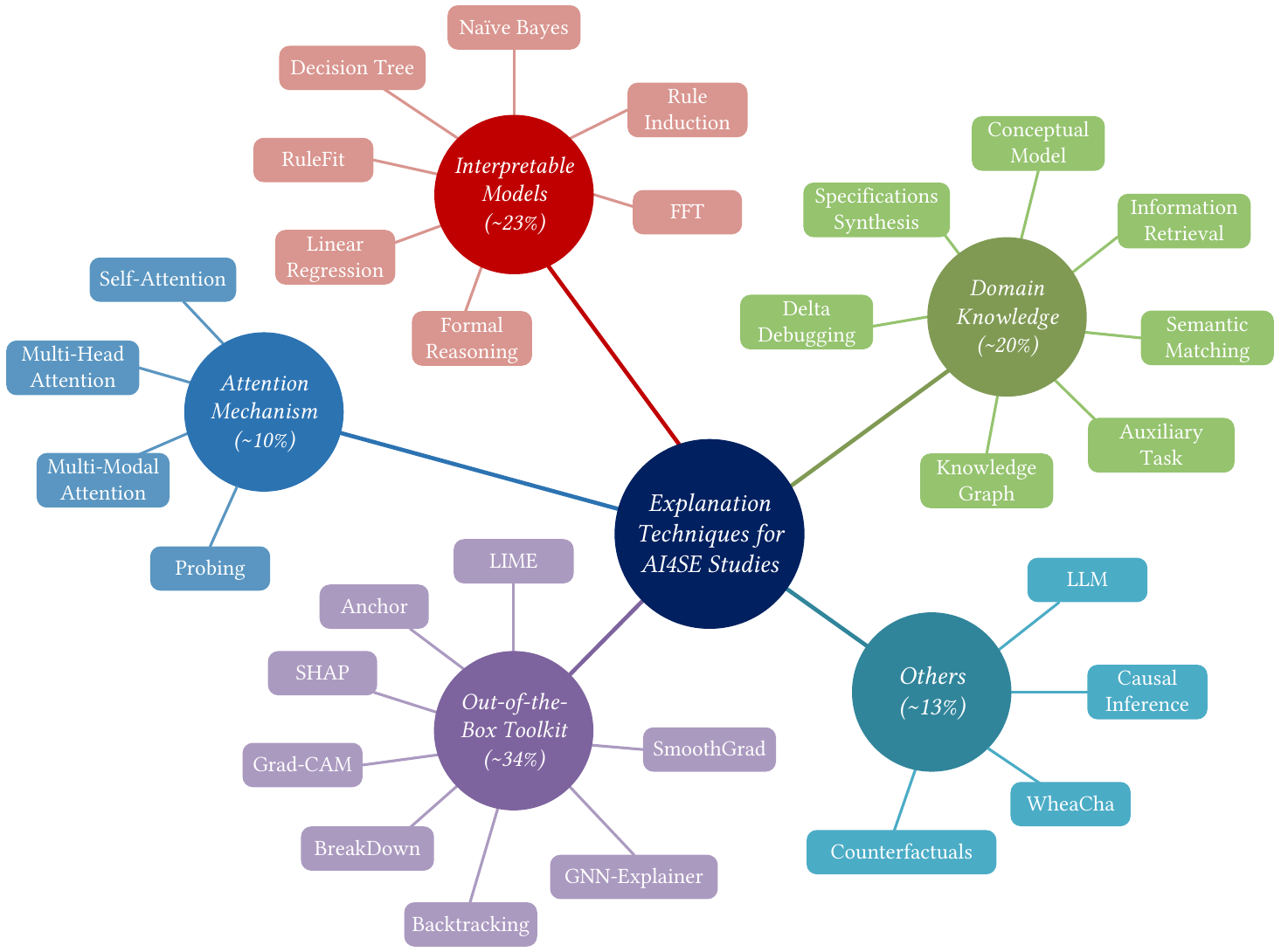}
\caption{XAI technique taxonomy \& distribution.}
\label{ExpTec}
\Description{}
\end{figure}

We first discuss various explanation approaches employed by existing XAI4SE studies. One classical practice is building a taxonomy of XAI techniques used in our surveyed literature. However, we note that the XAI community lacks a formal consensus on the taxonomy, as the landscape of explainability is too broad, involving substantial theories related to philosophy, social science, and cognitive science \cite{speith2022review}. In addition, these taxonomies are mostly developed for general purpose or specific downstream applications such as healthcare \cite{XAI4Medical2} and finance \cite{DBLP:journals/corr/abs-2309-11960}, and may not be applicable to the SE field. As a countermeasure, we summarize the XAI techniques used in primary studies, and propose a novel taxonomy applicable to the field of SE. In particular, from an integration perspective, most XAI techniques studied in this review can be categorized into five groups: \emph{Out-of-the-Box Toolkit (OT)} ($\approx$34\%), \emph{Interpretable Model (IM)} ($\approx$23\%), \emph{Domain Knowledge (DK)} ($\approx$20\%), \emph{Attention Mechanism (AM)} ($\approx$10\%), as well as a set of other custom, highly tailored approaches ($\approx$13\%). Figure \ref{ExpTec} illustrates the various types of XAI techniques that we extracted from our selected studies.

\noindent\textbf{Out-of-the-Box Toolkit (OT).}
Embracing off-the-shelf techniques from the field of XAI served as a natural starting point for researchers given the surge in publications and widespread adoption across various industries. Examining the prevalence of various different types of XAI tools, we found that \textbf{Feature Perturbation} \cite{LIME,SHAP,DBLP:conf/nips/YingBYZL19} are the most popular approaches, followed by gradient-based \cite{DBLP:conf/icml/SundararajanTY17,DBLP:journals/ijcv/SelvarajuCDVPB20} and decomposition-based approaches \cite{LRP,DBLP:journals/pr/MontavonLBSM17,DBLP:conf/icml/ShrikumarGK17}. The prevalence of perturbation-based approaches is expected, as they can work at various levels including embeddings vectors \cite{DBLP:conf/sigsoft/Xiao0XZ023}, source code \cite{DBLP:journals/tse/Wattanakriengkrai22}, texts \cite{DBLP:journals/ese/SchulteLH24}, and data structure \cite{IVDETECT}, which are common types of artifacts being used in AI4SE approaches. An early representative perturbation-based approach is \textbf{Local Interpretable Model-agnostic Explanations (LIME)} \cite{LIME}. Specifically, LIME first perturbs the to-be-explained instance in the high-dimensional feature space to randomly generate synthetic neighbors. Then, based on their prediction results derived from the global black-box model, LIME trains a local surrogate model (e.g., decision tree, linear regression) to produce an explanation. LIME has been successfully applied to various SE tasks, such as defect prediction \cite{DBLP:journals/jss/ZhengSCD22,DBLP:conf/wcre/LeeL23,DBLP:journals/tse/Wattanakriengkrai22}, OSS Sustainability Prediction \cite{DBLP:conf/sigsoft/Xiao0XZ023}, and test case generation \cite{DBLP:conf/issre/AdigunHCF23}. \textbf{SHapley Additive exPlanations (SHAP)} \cite{SHAP}, which stems from game theory, is another popular XAI technique based on perturbation-based feature attribution. It assigns each feature an fair value, otherwise known as \emph{shapley value}, to measure its contribution to the model's output. Features with positive SHAP values positively impact the prediction, and vice versa. Similar to LIME, SHAP is also model-agnostic, thus it can be used to explain any ML model. For example, Widyasari et al. \cite{DBLP:conf/iwpc/WidyasariPHTZ022} applied a tree ensemble model-specific variant, TreeSHAP \cite{TreeSHAP}, to identify which code statements are important in each failed test case. However, post-hoc approaches such as LIME and SHAP are computationally expensive. As a consequent, they are usually limited to simpler problems with a small number of features.

%Feature attribution-based explanations aim to measure the relevance (i.e., attribution score) of each input feature to a model’s prediction by ranking the explanatory score. The score could range from a positive value that shows its contribution to the model’s prediction, to a zero that would mean the feature has no contribution, to a negative value which means that removing that feature would increase the probability of the predicted class. They can be broadly divided into (\ding{182}) perturbation-based approaches that make feature perturbations while analyzing prediction change, e.g., LIME \cite{LIME}, GNNExplainer \cite{DBLP:conf/nips/YingBYZL19}, and SHAP \cite{SHAP}, (\ding{183}) gradient-based approaches that propagate importance signals backward through all neurons of the network, e.g., Integrated Gradients (IG) \cite{DBLP:conf/icml/SundararajanTY17} and Grad-CAM \cite{DBLP:journals/ijcv/SelvarajuCDVPB20}, and (\ding{184}) decomposition-based approaches that break down the relevance score into linear contributions from the input, e.g. Layer-wise Relevance Propagation (LRP) \cite{LRP}, Deep Taylor Decomposition (DTD) \cite{DBLP:journals/pr/MontavonLBSM17}, and DeepLIFT \cite{DBLP:conf/icml/ShrikumarGK17}.

% \begin{figure}
%   \centering
%   \includegraphics[width=.6\linewidth]{Figure/IM.pdf}
% \caption{Sample decision tree used for explainable re-opened bug prediction.}
% \label{IMTree}
% \Description{}
% \end{figure}

\begin{wrapfigure}{r}{.6\textwidth}
    \vspace{-5mm}
    \centering
    \includegraphics[width=.6\textwidth]{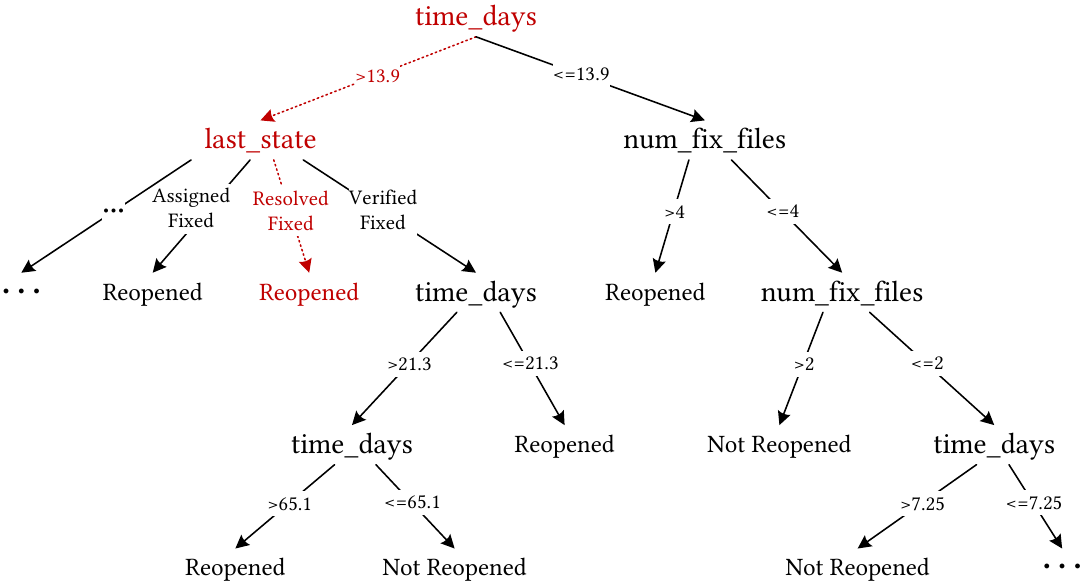}
    \caption{Sample decision tree used for explainable re-opened bug prediction.}\label{IMTree}
    \vspace{-5mm}
\end{wrapfigure}

\noindent\textbf{Interpretable Model (IM).}
As pointed out by Chen et al. \cite{DBLP:conf/sigsoft/ChenFKM18}, complex models did not always perform better than simpler alternatives. Thus, for certain SE tasks, the easiest way to achieve explainability is to construct interpretable models, such as Decision Tree (DT), Linear Regression (LR), and Naïve Bayes (NB). These models have built-in explainability by nature. For instance, DT predicts the value of a target variable by learning simple \code{if-then-else} rules inferred from the data features. The tree structure is ideal for capturing interactions between features in the data, and also has a natural visualization of a decision making process. Taking Figure \ref{IMTree} as an example, each node in a decision tree may refer to an explanation, e.g. when the \code{time\_days} variable (i.e., the number of days to fix the bug) is greater than 13.9 and the last status is \code{Resolved Fixed}, then the bug will be re-opened \cite{DBLP:journals/ese/ShihabIKIOAHM13}. In addition, interpretable models can serve as post-hoc surrogates to explain individual predictions of black-box models. The goal behind this insight is to leverage a relatively simpler and transparent model to approximate the predictions of the complicated model as best as possible, and at the same time, provide explainability. Surrogate models have shown effectiveness in explaining AI4SE approaches built upon more complex ML/DL models such as deep neural networks. Examples include vulnerability detection \cite{DBLP:journals/tosem/ZouZXLJY21}, defect prediction \cite{Pyexplainer}, and program repair \cite{DBLP:conf/msr/MarkovtsevLMSB19}.

\noindent\textbf{Domain Knowledge (DK).}
So far, we have encountered several XAI techniques to explain black-box models. However, there are some limitations regarding them in practice. On the one hand, due to the weak capability of processing complex data, intrinsically interpretable models are prone to getting trapped into the trade-off dilemma between performance and explainability, i.e., sacrificing predictive accuracy in exchange for explainability. On the other hand, important features highlighted by out-of-the-box tools are not necessarily user-friendly in terms of understandability and usability. For example, the importance of a single token in a code snippet may not convey a sufficiently meaningful explanation. To address these challenges, some works take special steps to incorporate additional knowledge from experts into their explanations. Concretely, we identify two incipient trends in the application of domain knowledge: (\ding{182}) designing one or more auxiliary tasks related to the main task to provide additional insights regarding the input data, and (\ding{183}) the use of an external knowledge database curated by experts. For example, to explain why a program/commit was predicted as vulnerable, recent works proposed to predict vulnerability types \cite{VulExplainer}, identify key aspects \cite{DBLP:conf/icse/SunXLXZHZ23}, generate vulnerability descriptions \cite{VulTeller,DBLP:conf/icse/MahbubSR23}, search for similar issues \cite{SVulD}, etc. To assist developers in understanding the return results of neural code search tools, XCoS \cite{10.1145/3593800} constructed a background knowledge graph, and regarded it as an external knowledge base to provide conceptual association paths, relevant descriptions, and additional suggestions, as explanations.

\noindent\textbf{Attention Mechanism (AM).}
As an increasingly common ingredient of neural architectures, attention mechanism has been widely applied to various SE tasks. Besides providing substantial performance benefits, it allows users to understand which parts of an input a model is most interested in through assigned weights, making the use of attention an intuitive option in practice. For example, Li et al. \cite{DBLP:conf/wcre/LiCZSP22} employed an attention-based GNN model, named GAT \cite{GAT}, to weigh the importance of neighboring code graph nodes in runtime exception detection. These computed weights were then used to visualize the importance of various edges to the detected runtime exception. Similarly, Wang et al. \cite{DBLP:conf/kbse/WangLLCLW20} visualized the feature weight of each word with the help of single- and multi-head attention mechanism to offer word- and phase-level explainability for why a comment was classified as self-admitted TD. Apart from explaining a model's individual predictions, attention mechanism can also offer insights into the inner workings of foundation models, providing a powerful tool for software engineering in the era of \textbf{Large Language Models (LLMs)}. A representative technique is probing, which trains a shallow classifier on top of the pre-trained or fine-tuned LLMs to identify certain knowledge/linguistic properties acquired by the model. For instance, Ma et al. \cite{weima} designed four probing tasks to analyze the capabilities of code models in understanding syntax and semantics by directly recovering the syntax and semantic structures from the code representation.

Although attention is a core component of Transformers, the effectiveness of attention-based explanations remains a topic of ongoing debate. There is skepticism about whether the attention mechanisms, which require substantial retraining, evaluation, and validation, can significantly enhance explainability in XAI4SE. Recent studies \cite{DBLP:conf/kbse/PaltenghiP21,DBLP:journals/pacmse/KouCW0024} have highlighted a persistent misalignment between human attention and model-generated attention in all code-based LLMs, emphasizing the need for neural models that more closely mimic human reasoning.

\noindent\textbf{Others.}
Apart from the above approaches, we also observed a number of XAI techniques, such as \emph{Causal Inference} and LLMs, that are specifically tailored for given SE tasks or neural architectures. Theory of Causation \cite{DBLP:conf/icse/VelascoPRP24} endows the model with the ability to pursue real causality without the interference from confounding factors. Such mechanism is a useful verification tool to achieve a more complete understanding of black-box neural models in SE tasks. For example, Palacio et al. \cite{DBLP:journals/tse/NaderPalacioVCRMP24} proposed a post-hoc approach specific to neural code models that provides programming language-oriented explanations based upon causal inference. To further provide actionable advice, some works employed counterfactuals-based causal inference paradigm to understand how the model reacts to feature changes. Cito et al. \cite{DBLP:conf/icse/CitoDMC22} proposed a Masked Language Modeling (MLM)-based perturbation algorithm, which replaces each token with a blank ``mask'' and uses MLM to come up with a plausible replacement for the original token, for generating counterfactual explanations. Counterfactuals-based explanation not only reveals which region of the input program is used by the code model for prediction, but also conveys critical information on how to modify the code so that the model will change its prediction. Recently, inspired by the remarkable performance in natural language understanding and logic reasoning, LLMs have been (in)directly integrated into the workflow of AI4SE approaches to offer explainability \cite{DBLP:conf/iwpc/0001TM0MCZYJ24,DBLP:conf/wcre/WidyasariANS024}. For instance, given an identified TODO-missed commit, Wang et al. \cite{TODO} fed it with the most similar historical commit to the LLM module to analyze and suggest \emph{where} and \emph{what} comment users should insert.

\subsubsection{Exploratory Data Analysis.}
% \begin{figure}[t]
%   \centering
%   \includegraphics[width=.9\linewidth]{Figure/RQ2A1.pdf}
% \caption{XAI techniques by SE task.}
% \label{XAIBT}
% \Description{}
% \end{figure}

%Fig. \ref{XAIBT} delineates the application status of different types of explanation approaches in \textcolor{red}{23} previously summarized SE tasks. 
Overall, \emph{Out-of-the-Box Toolkit} is the most prevalent explainability technique in XAI4SE research, followed by \emph{Interpretable Model}, \emph{Domain Knowledge}, and \emph{Attention Mechanism}. Given the rapid development of the XAI community, the popularity of out-of-the-box toolkit and attention mechanism is not surprising. Meanwhile, the prevalence of interpretable model is also logical, as they are inherently more understandable to humans and can serve as surrogate models to explain individual predictions of complex deep neural networks, which are more common in AI4SE approaches. Due to the diversity of SE data, domain knowledge-based explanations are also popular. The explanations are intended to give control back to the user by helping them understand the model and offering additional insights into the input data.

In addition to the prevalence, selecting the most suitable explanation approach for a specific task is also a crucial aspect that needs to be carefully considered. We further extracted the selection rationale for XAI techniques from the primary studies, and classified them into the following three categories:

\noindent\textbf{Task Fitness.}
Given the inherent differences in feature engineering and functional requirements among various AI4SE workflows, some studies selected XAI techniques based on their characteristics and fitness with target SE tasks. For instance, the explanations for most of the feature engineering-based AI4SE pipelines (e.g., defect prediction \cite{Pyexplainer} and OSS sustainability prediction \cite{DBLP:conf/sigsoft/Xiao0XZ023}) were derived by using interpretable models, such as decision tree or RuleFit \cite{RuleFit}, thanks to their strong ability to analyze and extract human-understandable rules from hand-crafted feature metrics which are usually limited.

\noindent\textbf{Model Compatibility.}
Since certain crafted XAI techniques have strict application scenarios, e.g., requiring the internal architecture or parameter information of to-be-explained models, some researchers determined the most suitable XAI techniques from those compatible with their employed models \cite{IVDETECT,DBLP:conf/kbse/0007LY0CWM22,DBLP:conf/issta/HuWLPWZ023}. For instance, due to the great performance of deconvolution \cite{DBLP:conf/eccv/ZeilerF14} in providing visual explanations for CNN-based applications, Ren et al. \cite{DBLP:journals/tosem/RenXXLWG19} leveraged a targeted backtracking technique to extract prominent phrases that contribute most to the decision whether the comment is a SATD or not from the input comment as explanations.

\noindent\textbf{Stakeholder Preference.}
In addition to the task fitness and model compatibility, stakeholder preference is also one of the important factors affecting the selection of XAI techniques. Generally speaking, model users aim to utilize XAI techniques to better understand the output of a deployed model and make an informed decision, while model designers focus on using XAI techniques during model training and validation to verify that the model works as intended. In addition, the explanations would be generated for distinct purposes at different levels of expertise even when considering a single stakeholder. For instance, to assist software developers in understanding a defective commit, some approaches simply highlight the lines of code that the model thinks are defective \cite{DBLP:journals/tse/Wattanakriengkrai22}, while others extract human-understandable rules \cite{Pyexplainer}, or even natural language descriptions \cite{DBLP:conf/icse/MahbubSR23} from the defective code that can serve as actionable and reusable patterns or knowledge.

%Among all papers reviewed, Wang et al. \cite{10.1145/3593800} leveraged a structured conceptual tree to display the results of query scoping, the identified conceptual association paths. Users can interact by selecting the anchor concept relevant to the query which will update the results of code snippets related to the selected concepts.

\subsection{RQ\texorpdfstring{$_{2b}$}:: What Format of Explanation Is Provided for Various SE Tasks?}\label{RQ2b}
\begin{figure}[t]
        \centering
        \subcaptionbox{Numerical explanation.}{
        \centering
        \includegraphics[width = .23\linewidth]{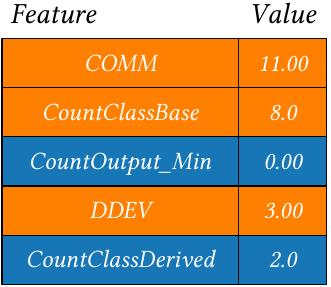}
    }
    \hspace{5mm}
        \subcaptionbox{Source Code explanation.}{
        \includegraphics[width = .35\linewidth]{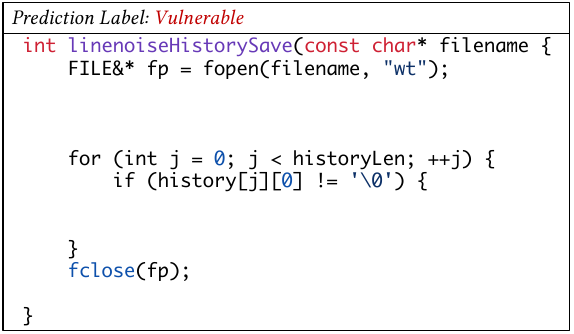}
    }
    \hspace{5mm}
        \subcaptionbox{Rule explanation.}{
        \includegraphics[width = .28\linewidth]{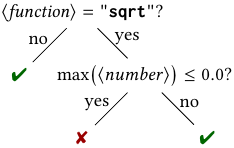}
    }
    \hspace{5mm}
        \subcaptionbox{Textual explanation.}{
        \includegraphics[width = .38\linewidth]{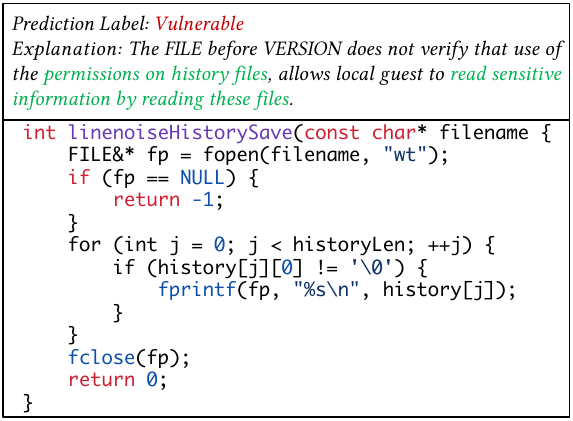}
    }
    \hspace{5mm}
        \subcaptionbox{Visual explanation.}{
        \includegraphics[width = .35\linewidth]{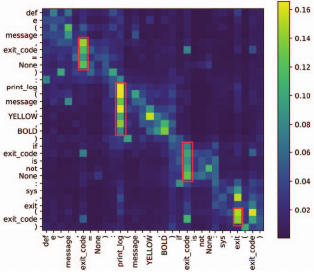}
    }
        \caption{Different formats of explanation.}
    \Description{}
    \label{VulTellerExp}
\end{figure}

Next, to analyze the formats of explanation being used in XAI4SE research, we provide a high-level classification, along with descriptive statistics, as to why some formats of explanation were used for particular SE tasks. In total, we observed five major explanation formats: \emph{Numeric}, \emph{Text}, \emph{Visualization}, \emph{Source Code}, and \emph{Rule}, as illustrated in Figure \ref{VulTellerExp}.

% \begin{figure}[t]
%   \centering
%   \includegraphics[width=\linewidth]{Figure/RQ2A2.pdf}
% \caption{Vulnerability explanation generated by VulTeller \cite{VulTeller}, which transfers detected vulnerable code to natural language descriptions conveying the reason for vulnerabilities.}
% \label{VulTellerExp}
% \Description{}
% \end{figure}

\noindent\textbf{Numeric.}
Numerical explanations, which are capable of conveying information in a compact format, focus on quantifying the positive or negative contribution of an input variable to the model prediction. Such importance scores can either be directly used as explanations \cite{DBLP:journals/tse/JiarpakdeeTDG22,DBLP:journals/jss/ZhengSCD22} or serve as indicators to guide key feature selection \cite{DBLP:conf/sigsoft/SunXTDLWZCN23}. Popular examples of numerical explanations are LIME \cite{LIME} and SHAP \cite{SHAP}. Esteves et al. \cite{DBLP:journals/ase/SantosFVVZ20} used SHAP values to understand the CK metrics \cite{CKMetrics} that influenced the defectiveness of the classes. In the same context, Lee et al. \cite{DBLP:conf/wcre/LeeL23} employed three widely-used model-agnostic techniques, including LIME, SHAP, and BreakDown \cite{DBLP:journals/rjour/StaniakB18}, to calculate the contribution of each feature for defect models' predictions. Xiao et al. \cite{DBLP:conf/sigsoft/Xiao0XZ023} leveraged the local explanations generated by LIME from the XGBoost model to analyze the contribution of variables to sustained activity in different project contexts. To reflect the global behavior of a complex JIT defect prediction model, Zheng et al. \cite{DBLP:journals/jss/ZhengSCD22} employed SP-LIME, a variant of LIME, to analyze the relationship between the features in the model and the final prediction results. SP-LIME explicitly selects representative and diverse (but not repetitive) samples, presenting a global view within the allocated budget of maximal features. Sun et al. \cite{DBLP:conf/sigsoft/SunXTDLWZCN23} used SHAP to guide the search for the feature that has largest malicious magnitude, i.e., having the potential to be manipulated by the adversary, to test the robustness of malware detectors.

\noindent\textbf{Text.}
In contrast to numerical explanations, textual natural language descriptions are easier to be comprehended by non-experts, offering clarity in understanding the behavior of intelligent SE models. Such textual explanations can either be derived from scratch by using generative models \cite{DBLP:conf/icse/MahbubSR23,VulTeller,DBLP:conf/icse/SunXLXZHZ23} or retrieved from external knowledge bases \cite{DBLP:journals/tosem/WuCGFLWL21,KGXQR}. For example, Zhang et al. \cite{VulTeller} leveraged a GRU-based decoder to generate vulnerability symptom- or reason-related descriptive sentences step by step. Such summarization-styled explanations can effectively bridge the cognitive gap between structured programming language and flatten natural language. Wu et al. \cite{DBLP:journals/tosem/WuCGFLWL21} built a semantic database based on malware key features and functional descriptions in developer documentation, and leveraged the mapping relation between the malicious behaviors and their corresponding semantics to generate reasonable descriptions that are easier for users to understand. Inspired by similar/homogeneous vulnerabilities that have similar root causes or lead to similar impacts, Ni et al. \cite{SVulD} first retrieved the most semantically similar problematic posts from SO and prioritized the most useful response based on a quality-first sorting strategy. Then, they employed the BERT-QA model \cite{BERT-QA} to extract the root cause, impact, and solution from the answers to the given questions as useful and understandable natural language explanations. Compared to generative models pre-trained on the human-labeled corpus, external knowledge bases such as Stack Overflow and Wikipedia\footnote{\url{https://www.wikipedia.org/}} offer structured knowledge models that explicitly store rich factual knowledge. Thus, they are well-known for their symbolic reasoning ability, which generates explainable results, and avoids hallucinations originating from generated statements that are factually incorrect.

\noindent\textbf{Visualization.}
% \begin{figure}[t]
%         \centering
%         \subcaptionbox{Heapmap}{
%         \centering
%         \includegraphics[width = .4\linewidth]{Figure/HeatMap.pdf}
%     }
%     \hspace{15mm}
%         \subcaptionbox{Bipartite Graph}{
%         \includegraphics[width = .4\linewidth]{Figure/Bipartite Graph.pdf}
%     }
%         \caption{\label{VisualMap}Visual explanations provided by different techniques. Adapted from \cite{DBLP:conf/icse/WanZZSXJ22}.}
% \Description{}
% \end{figure}
Besides explaining through numerical importance scores and textual natural language descriptions, users can understand the behavior of the underlying model through the form of visuals. Humans, in general, can process visual information faster and much easier as compared to other information \cite{munzner2014visualization}. Common techniques involve visualizing attention heads for a single input using bipartite graphs or heatmaps. They are simply disparate visual representation of attentions, one as a graph and the other as a matrix. Wang et al. \cite{DBLP:conf/kbse/WangLLCLW20} visualized the attention weights of the most important words and phrases that have contributed to the model's predictions. In addition, some approaches developed interactive User Interface (UI) to provide visual explanations \cite{DBLP:journals/tse/JiangSWCYZ23,10.1145/3593800}. For instance, Jiang et al. \cite{DBLP:journals/tse/JiangSWCYZ23} designed several contribution mining algorithms to infer the key elements in code that contribute to the generation of the key phrases in the comments. When a developer intends to comprehend the code, the UI loads the auto-generated comments and presents to the developers the graphic illustration by coloring the important phrases and the corresponding parts in the source code. In this way, the developer can check whether the auto-comments correctly describe the intention of the code.

\noindent\textbf{Source Code.}
Some attempts borrowed certain classical techniques, such as program mutation \cite{DBLP:journals/tsmc/SrinivasP94} and \textbf{Delta Debugging (DD)} \cite{DBLP:journals/tse/ZellerH02}, from the field of software testing to search for important code snippets positively contributing to the model predictions. For instance, Geng et al. \cite{DBLP:conf/iwpc/GengWDWCZJ23} identified important code tokens contributing to the generation of a specific part of the summarization by checking which meaningful words disappeared in any of the summarization newly generated from mutants. Rabin et al. \cite{DBLP:conf/sigsoft/RabinHA21} leveraged DD to simplify a piece of code into the minimal fragment without reversing it original prediction label. The reduction process continues until the input data is either fully reduced (to its minimal components, depending on the task) or any further reduction would corrupt the prediction. In contrast to debugging-based techniques that can be applied to any DL architecture, Li et al. \cite{IVDETECT} employed a GNN-specific explanation framework, GNNExplainer \cite{DBLP:conf/nips/YingBYZL19}, to simplify the target code instance to a minimal statements subset.

%consisting of crucial statements while retaining the initial model prediction. 

%Along the lines of architecture-specific simplification, Hu et al. \cite{DBLP:conf/issta/HuWLPWZ023} investigated six famous GNN explainers to evaluate their explanation performance on vulnerability detectors from three different perspectives, which are effectiveness, stability, and robustness. Their empirical study showed that the explanation results provided by different explainers for vulnerability detection vary significantly, and the performance of all explainers was still not satisfactory.

%As source code is one of the primary data types that AI4SE models attempt to learn from, it serves as a natural format of explanation. 
%Wang et al. \cite{WheCha} further classified an entire input into two types of features: defining features (i.e., \emph{wheat}) representing the reasons models predict a specific label, and the rest (i.e., \emph{chaff}). They proposed a coarse-to-fine approach, named \emph{``Reduce and Mutate''}, to identify \emph{wheat} that code models use for prediction. 

\noindent\textbf{Rule.}
Rule, which can be organized in the form of \code{IF-THEN-ELSE} statements with \code{AND/OR} operators, is a schematic and logic format. Despite its complexity compared to visualization and natural language description, rule-based explanation is still intuitive for humans and useful for expressing combinations of input features and their activation values. Generally, these rules approximate a black-box model but have higher interpretability. Zou et al. \cite{DBLP:journals/tosem/ZouZXLJY21} identified important code tokens whose perturbations lead to the variant examples having a significant impact on the prediction of the target model via heuristic searching, and trained a decision tree-based regression model to extract human-understandable rules for explaining why a particular example is predicted into a particular label. To explain the individual predictions of the black-box global model, Pornprasit et al. \cite{Pyexplainer} built a RuleFit-based local surrogate model, which combined the strengths of decision tree and linear model, to understand the logical reasons learned from the rule features. Cito et al. \cite{DBLP:conf/sigsoft/CitoD0M021} proposed a rule induction technique which produced decision lists based on features and mispredicted instances to explain the reasons for mispredictions.

\begin{figure}[t]
  \centering
  \includegraphics[width=.85\linewidth]{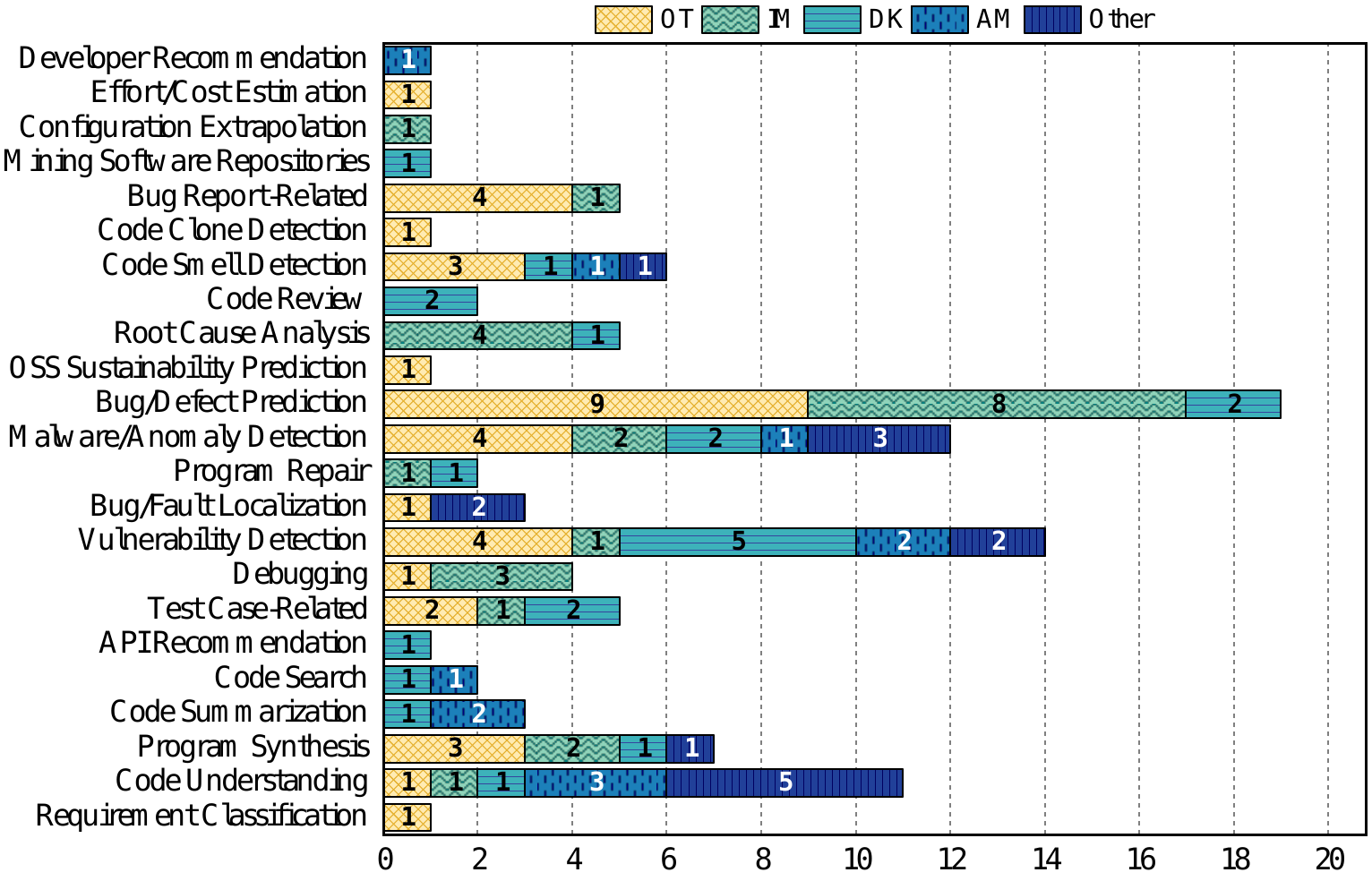}
\caption{Explanation formats by SE task \& XAI techniques taken.}
\label{OutputBT}
\Description{}
\end{figure}

\subsubsection{Exploratory Data Analysis}
Figure \ref{OutputBT} shows a breakdown of the relationships between the SE tasks, the explanation techniques taken, and explanation formats. We found that the distribution of different explanation formats is relatively even except \emph{Visualization} ($\approx$9\%), and the most common format being used is \emph{Numeric} ($\approx$27\%). The primary reason is that numerical features are a common source of information across all aspects of data-driven methodologies. SE tasks such as \emph{Bug/Defect Prediction} and \emph{Code Smell Detection} commonly use a collection of hand-crafted numerical features (e.g., code metrics \cite{CKMetrics}, smells \cite{DBLP:conf/msr/PecorelliPKL20}, permission and API calls \cite{DBLP:conf/ictai/PeiravianZ13}) to train a model. Measuring the relevance of each input feature to a model’s prediction is intuitive and has been well established within the field of XAI, hence it is not surprising that the majority of reviewed studies focus on this format. We also noticed that visual explanation is less commonly used, only accounting for a total of 10 primary studies. One possible reason contributing to its relative lack of adoption lies in that, although visualization can provide a fast and straightforward explanation for practitioners (e.g., a developer, domain expert, or end-user) who are inexperienced in ML/DL \cite{DBLP:journals/tvcg/SpinnerSSE20}, it can only convey limited information and requires post-processing for further use.

In addition to the prevalence, we observed that the formats of explanation varied even in a single SE task. As an example, the explanations generated for binary vulnerability detectors in our surveyed studies can be text (e.g., vulnerability descriptions \cite{VulTeller,SVulD}, types \cite{CoLeFunDa,DBLP:conf/icse/SunXLXZHZ23}), source code (e.g., code statements \cite{DBLP:conf/sigsoft/SunejaZZLM21,IVDETECT}), or rules \cite{DBLP:journals/tosem/ZouZXLJY21}. This diversity helps to satisfy the personalized needs of the stakeholders who have different intents and expertise. 

%We expect that this trend will persist, with the continued proposal and application of new XAI techniques to various SE tasks.

%The prevalence of source code-based explanation is not surprising, given the rise in the number of publicly available code repositories, a series of AI models \cite{DBLP:conf/acl/GuoLDW0022,DBLP:conf/iclr/GuoRLFT0ZDSFTDC21} have achieved notable advancements in various downstream code-centric SE tasks, such as code generation, understanding, and analysis \cite{CodeIntelligent}, making it a natural way to obtain human-understandable explanations. 

%describes the proportion of various formats of explanations we summarized from \textcolor{red}{108} primary studies in detail and gives examples.

%Fig. \ref{OutputBT} provides an overview of the distribution of explanation formats adopted by different SE tasks. 

\begin{myhbox}{\ding{45} $\blacktriangleright$ RQ$_{2}$ - Summary $\blacktriangleleft$}
\begin{itemize}[leftmargin=2em]
\item Our exploratory data analysis revealed five commonly used XAI techniques, including \emph{Out-of-the-Box Toolkit (OT)} ($\approx$34\%), \emph{Interpretable Model (IM)} ($\approx$23\%), \emph{Domain Knowledge (DK)} ($\approx$20\%), \emph{Attention Mechanism (AM)} ($\approx$10\%), as well as a subset of other custom, highly tailored approaches ($\approx$13\%).
%Among them, \emph{OT} is by far the most popular option in our surveyed studies.
\item We summarized the selection strategies for XAI techniques in SE tasks into three main categories: \emph{Task Fitness}, \emph{Model Compatibility}, and \emph{Stakeholder Preference}.
\item A variety of explanation formats have been explored in our surveyed studies, with the main formats utilized being \emph{numeric} ($\approx$27\%), \emph{text} ($\approx$23\%), \emph{visualization} ($\approx$9\%), \emph{source code} ($\approx$20\%), and \emph{rule} ($\approx$20\%).
\item  We found a strong correlation between the SE tasks, the explanation techniques taken, and explanation formats. Additionally, the formats of explanation also varied even in a single SE task. This diversity helps to satisfy the personalized needs of the stakeholders who have different intents and expertise.
\end{itemize}
\end{myhbox}

%\item \textcolor{red}{Numeric} is the most prevalent explanation format in XAI4SE studies. Few studies employed visualization as their preferred format of explanation due to the limited information it conveyed.

\section{RQ\texorpdfstring{$_3$}:: How well do XAI techniques perform in supporting various SE tasks?}\label{RQ3}
Evaluating the effectiveness of proposed solutions against existing datasets and employing baseline comparisons is a standard practice in AI4SE research. In this RQ, we endeavor to investigate the influence of XAI4SE research by scrutinizing the effectiveness of the techniques proposed in the studies under consideration. Our analysis primarily focuses on evaluating metrics on a task-specific basis, aiming to encapsulate the prevailing benchmarks and baselines within the field of XAI4SE research.

\subsection{RQ\texorpdfstring{$_{3a}$}:: What Baseline Techniques Are Used to Evaluate XAI4SE Approaches?}
\begin{figure}
        \centering
        \subcaptionbox{Distribution of baseline usage.\label{RQ3A}}{
        \centering
        \includegraphics[width = .3\linewidth]{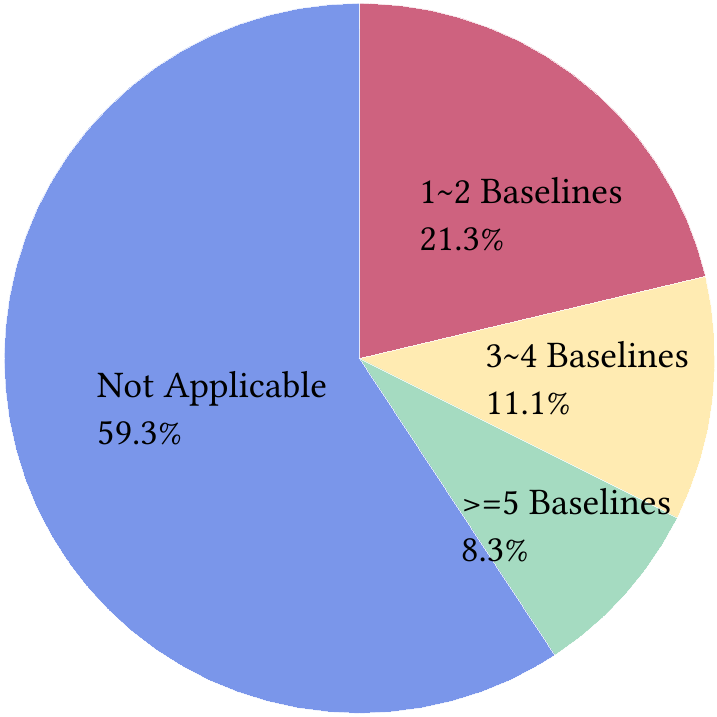}
    }
    \hspace{15mm}
        \subcaptionbox{Approach availability.\label{RQ3B}}{
        \includegraphics[width = .3\linewidth]{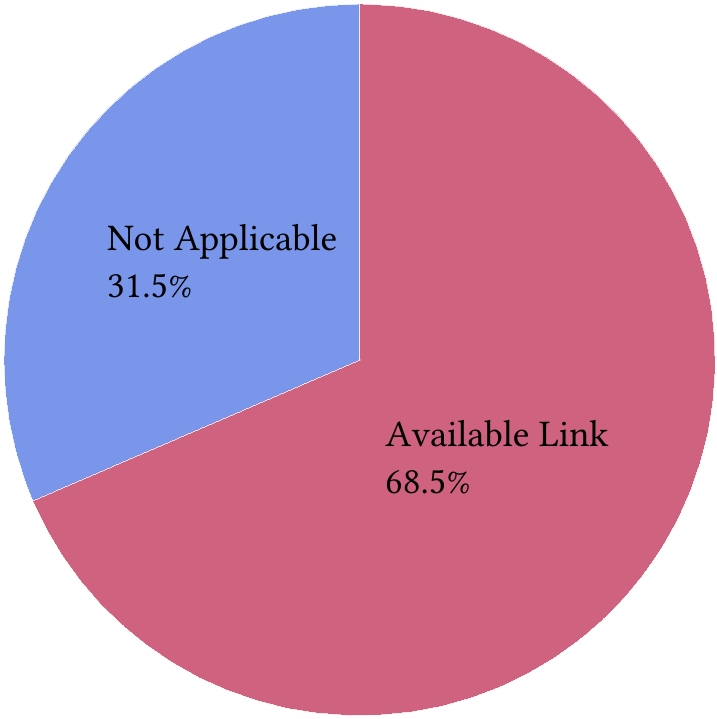}
    }
        \caption{Statistical information of baseline techniques and XAI4SE approaches.}
    \Description{}
    \end{figure}
For RQ$_{3a}$, we scrutinize the baseline techniques employed for comparative analysis in our selected primary studies. Although common baselines for particular software engineering tasks were identified, it was noted that a substantial portion of the literature autonomously developed unique baselines. The extensive variety and volume of these baselines precluded their detailed inclusion within the body of this manuscript. Therefore, we included the listing of baselines that each paper compared against on our interactive website at \url{https://riss-vul.github.io/xai4se-paper/}. Figure \ref{RQ3A} briefly analyzes the distribution of baseline usage in XAI4SE studies. Approximately $\approx$60\% of the studies reviewed do not engage in comparisons with any baseline, whereas a minority contrasts their findings with more than four distinct methods ($\approx$8.3\%). It was observed that numerous baseline techniques consist of established white-box models with transparent algorithms or conventional expert systems. This trend may be attributed, in part, to the nascent stage of XAI4SE research, which has resulted in a limited range of existing XAI-centric comparatives. As XAI4SE progresses towards maturity, an evolution towards evaluations incorporating benchmarks against established XAI-centric methodologies is anticipated. Furthermore, it was noted that the selection of baseline techniques exhibits a high degree of specificity, varying significantly even among studies addressing identical software engineering tasks. For example, to evaluate the effectiveness of their proposed explainable vulnerability detection approaches, eight out of 14 primary studies employed at least two baselines for evaluation, while only two papers \cite{DBLP:conf/issta/HuWLPWZ023,DBLP:conf/issta/Chu00W0S0024} overlap slightly in terms of the baselines.

A concerning trend identified in our review is the lack of publicly accessible implementations for many XAI4SE approaches. We conducted a manual inspection of all links provided within each study. In instances where a link to a replication package was available, we assessed its contents for source code and relevant documentation. Absent any direct links, we also endeavored to locate either the original replication package or an equivalent reproduction package on GitHub using the title of the paper as a search query. As depicted in the pie chart in Figure \ref{RQ3B}, approximately only 68\% of the primary studies offer accessible replication packages. Among the remaining studies with replication packages, a considerable portion of them propose XAI techniques for specific SE tasks for the first time, such as \emph{Mining Software Repositories} \cite{KGXQR} and \emph{OSS Sustainability Prediction} \cite{DBLP:conf/sigsoft/Xiao0XZ023}. This, in part, explains the proliferation of highly individualized baseline approaches. Researchers often lack access to common baselines for comparison, compelling them to implement their own versions. The robustness of results in such papers may be compromised, as many do not provide information about the baselines used. Moreover, distinct implementations of the same baselines could lead to confounding results when assessing purported improvements. While we anticipate that the set of existing publicly available baselines will improve over time, we also recognize the necessity for well-documented and publicly available baselines, accompanied by guidelines that dictate their proper dissemination.

\subsection{RQ\texorpdfstring{$_{3b}$}:: What Benchmarks Are Used for These Comparisons?}
% \begin{figure}[t]
%   \centering
%   \includegraphics[width=.33\linewidth]{Figure/RQ3A.pdf}
% \caption{Construction strategies of benchmarks.\label{RQ3C}}
% \label{Bench4Task1}
% \end{figure}
\begin{figure}[t]
  \centering
  \includegraphics[width=\linewidth]{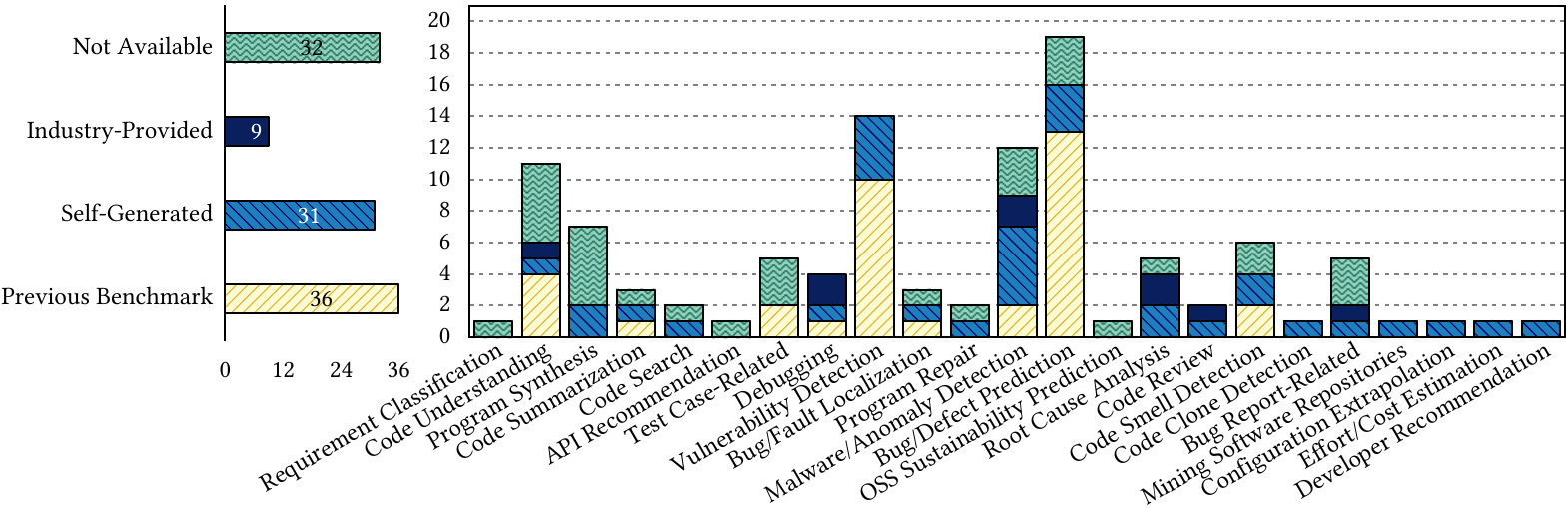}
\caption{Collection strategies of benchmarks by SE task.}
\label{Bench4Task2}
\Description{}
\end{figure}
For RQ$_{3b}$, we investigated the collection strategies of benchmarks in XAI4SE studies. As can be seen from the data in Figure \ref{Bench4Task2} (left), only 36 ($\approx$33\%) studies used previously curated benchmarks for evaluating XAI4SE approaches. The selection of open-source benchmarks is often motivated by their compelling nature in assessing the performance of AI-driven methodologies, which facilitates the reproducibility and replication by subsequent studies. Given the nascent emergence of XAI4SE research, there is an observed scarcity of appropriate benchmark datasets. This also explains why there is a considerable amount (28.7\%) of self-generated benchmarks. This trend within XAI4SE is worrying, as there are few instances where XAI approaches can appropriately compare against one another with available benchmarks. In our online repository, we recorded the accessible benchmark links provided by primary studies. Our aim is to assist researchers by offering insights into the available benchmarks for evaluating methodologies within distinct software engineering tasks. Furthermore, we advocate for future scholars to share their self-created benchmarks publicly, thereby furnishing a valuable resource that facilitates not only comparative analyses among different methodologies but also broadens the dataset accessible for Explainable AI techniques.

While the adoption of pre-existing benchmarks was infrequent across our surveyed studies, we did observe a subset of benchmarks that recurred within our primary studies. A comprehensive delineation of the types of the benchmarks employed in these primary studies is depicted in Figure \ref{Bench4Task2} (right). For vulnerability detection, we found that the Big-Vul \cite{FAN} dataset was used frequently, including evaluating the accuracy of the key aspects extracted from the detected vulnerabilities (e.g., vulnerable statements \cite{DBLP:conf/issta/HuWLPWZ023,IVDETECT}, vulnerability types \cite{VulExplainer}). Additionally, the dataset released by Yatish et al. \cite{DBLP:conf/icse/YatishJTT19} was employed for benchmarking prior XAI techniques targeting defect prediction models \cite{DBLP:journals/tse/JiarpakdeeTDG22,DBLP:conf/wcre/LeeL23}.

%in the context of defect prediction, 

\subsection{RQ\texorpdfstring{$_{3c}$}:: What Evaluation Metrics Are Employed to Measure XAI4SE Approaches?}\label{SectionRQ3c}
% \begin{figure}[t]
%   \centering
%   \includegraphics[width=.3\linewidth]{Figure/RQ3C1.pdf}
% \caption{\textcolor{red}{Evaluation strategies of XAI techniques.}}
% \label{MetricByStrategy}
% \Description{}
% \end{figure}

\begin{wrapfigure}{r}{.3\textwidth}
    \vspace{-3mm}
    \centering
    \includegraphics[width=.3\textwidth]{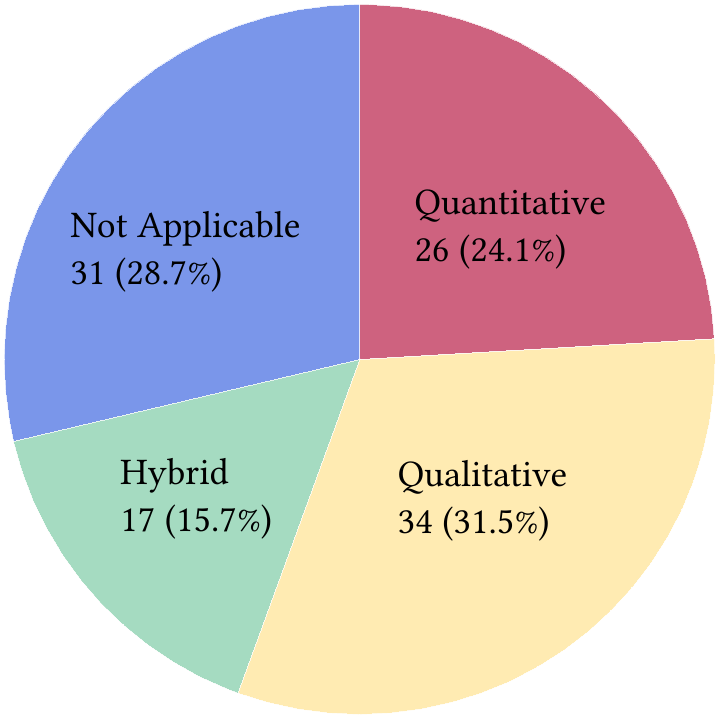}
    \caption{Evaluation strategies of XAI techniques.}\label{MetricByStrategy}
    \vspace{-3mm}
\end{wrapfigure}
With an increasing number of XAI techniques, the demand grows for suitable evaluation metrics \cite{XAI1,DBLP:journals/jair/BurkartH21,DBLP:journals/csur/GuidottiMRTGP19}. This need is not only recognized by the AI community, but also by the SE community as evaluating the performance of XAI techniques is a crucial aspect of their development and deployment \cite{DBLP:journals/corr/abs-2308-12415}. Figure \ref{MetricByStrategy} describes the distribution of evaluation strategies found in this SLR. The absence of objective, quantifiable evaluation is not surprising, given the scarcity of reliable benchmarks that are publicly available, as discussed above. In our analysis of utilized evaluation strategies within work on XAI4SE, we observed that nearly one-third ($\approx$32\%) of primary studies only adopted anecdotal evidence or user study, instead of quantitative metrics to evaluate their proposed approaches. That's to be expected because traditional performance metrics exist to evaluate prediction accuracy and computational complexity, while auxiliary criteria such as explainability may not be easily quantified \cite{doshi2018considerations}. Among 43 papers that performed quantitative evaluation, we found that the SE community also has yet to agree upon standardized evaluation metrics due to the absence of ground truths. Furthermore, due to the wide range of objectives associated with explainability in SE tasks, relying solely on a single evaluation metric may not adequately reflect the full spectrum of an XAI tool's performance. Consequently, researchers frequently utilize a variety of evaluation metrics, each designed to measure specific aspects of explainability. According to the best practice within the field of XAI \cite{10.1145/3583558}, these quantitative metrics can be clustered from a multi-dimensional view, named \textbf{Co-12 properties}. Table \ref{EvaluationMetric} describes each Co-12 property we identified, while listing the evaluation metrics that were mainly related with this property and the papers that applied these metrics. Since over 30 unique metrics are adopted by our collected primary studies, which is difficult to present them all in this SLR, we listed 16 main metrics across six major Co-12 properties, where each metric was researched by no less than two studies.

\begin{table*}[t]
 \caption{Metrics Used for Evaluation}
  \centering
  \scalebox{0.68}{
  \begin{tabular}{|l|lll|}
    \toprule
    \textbf{Property} & \textbf{Description} &\textbf{Metric} & References \\
    \midrule
    \multirow{6}*{Correctness}   &  \multirow{6}*{How faithful the explanation is w.r.t the black box.} & \%Consistency & \cite{Pyexplainer,DBLP:journals/jss/YangZZXZQ24} \\
    ~ & ~ & PCR & \cite{DBLP:journals/tosem/ZouZXLJY21,DBLP:journals/pacmse/Chen00024} \\
    ~ & ~ & Statistical Test  & \cite{Pyexplainer,DBLP:journals/jss/YangZZXZQ24,FOX} \\
    ~ & ~ & Distance-based & \cite{Pyexplainer,DBLP:journals/jss/YangZZXZQ24,FOX} \\
    ~ & ~ & R$^2$ & \cite{DBLP:journals/tse/JiarpakdeeTDG22,DBLP:journals/tse/LinTH22} \\
    ~ & ~ & RBO & \cite{FOX,DBLP:conf/wcre/LeeL23} \\
    \midrule
    Consistency & How deterministic the explanation approach is. & Statistical Test & \cite{DBLP:journals/jss/AwalR24,DBLP:journals/tosem/LyuRLCJ22,DBLP:journals/tse/LinTH22} \\
    \midrule
    Continuity & How continuous and generalizable the explanation function is. & Similarity-based & \cite{DBLP:conf/kbse/LiZZWLXCJ23,DBLP:conf/issta/HuWLPWZ023}\\
    \midrule
    Contrastivity & How discriminative the explanation is w.r.t. other events or targets.  & \%Unique & \cite{Pyexplainer,DBLP:journals/jss/YangZZXZQ24}\\
    \midrule
    \multirow{2}*{Compactness} & \multirow{2}*{The size of the explanation.} & Reduction Ratio & \cite{DBLP:conf/sigsoft/RabinHA21,WheCha} \\
    ~ & ~ & \#Rules & \cite{DBLP:conf/sigsoft/KampmannHSZ20,DBLP:conf/msr/MarkovtsevLMSB19,DBLP:journals/tse/PengM22}\\
    \midrule
    \multirow{4}*{Coherence} & \multirow{4}*{How accordant the explanation is with prior knowledge and beliefs.} & Alignment-based & \cite{DBLP:conf/icse/MahbubSR23,DBLP:conf/icse/SunXLXZHZ23,VulTeller,DBLP:conf/wcre/WidyasariANS024,DBLP:journals/pacmse/KouCW0024}\\
    ~ & ~ & Classification-based & \cite{VulExplainer,CoLeFunDa,DBLP:conf/esem/SarkerSWB23,DBLP:conf/icse/Cao0W0B0024,DBLP:conf/issta/Chu00W0S0024,DBLP:conf/issta/HuWLPWZ023,DBLP:conf/icse/GesiSGCA23,DBLP:conf/sigsoft/CitoD0M021} \\
    ~ & ~ & Ranking-based & \cite{IVDETECT,DBLP:conf/kbse/0007LY0CWM22,DBLP:journals/tse/Wattanakriengkrai22,DBLP:journals/tse/Wattanakriengkrai22}\\
    ~ & ~ & Statistical Test & \cite{DBLP:journals/tse/PaltenghiPHZ24,DBLP:conf/kbse/PaltenghiP21,DBLP:conf/apsec/AbidCJ23,DBLP:journals/tse/JiangSWCYZ23} \\
    \midrule
    Efficiency & The average runtime of generating one explanation per instance. & Runtime & \cite{FOX,WheCha,DBLP:conf/sigsoft/RabinHA21,DBLP:journals/pacmse/Chen00024}\\
    \bottomrule
  \end{tabular}}
  \label{EvaluationMetric}
\end{table*}

\noindent\textbf{Correctness.}
Correctness describes the degree at which an explanation technique approximates the behavior (either locally or globally) of the target model. In certain primary studies, it is also referred to as \textbf{Fidelity} \cite{DBLP:conf/kbse/LiZZWLXCJ23,DBLP:journals/tosem/ChengZWWBFGMW24}. Common metrics chosen to evaluate correctness include \emph{\%Consistency}, \emph{Positive Classification Rate (PCR)} \cite{DBLP:conf/ccs/GuoMXSWX18}, \emph{Statistical Test}, \emph{Distance-based}, $R^2$, and \emph{Ranked Biased Overlap (RBO)} \cite{RBO}. For instance, RBO is a similarity measure (assessed within the ranges of [0, 1]) that quantifies the differences between indefinite ranked lists. A higher value signifies stronger alignment or closeness between the explanations generated by different explainers. Given that numerical explanation (e.g., feature importance) is one of the most common explanation formats in our surveyed papers, the quantification of differences between feature importance is crucial for assessing model trustworthiness.

\noindent\textbf{Consistency.}
In addition to correct results, the generated explanations need to be consistent. That is, two models that give the same outputs for all inputs should have the same explanations in order to be useful for an expert. A common evaluation approach is statistical test, which computes the statistical differences among multiple runs on the same instance. For example, Awal et al. \cite{DBLP:journals/jss/AwalR24} assessed inconsistency by running the same explanation technique on 500 instance (100 consecutive iterations per instance) and measuring the rank difference of each metric. To measure whether the explanation results between two executions were statistically different, they applied the Wilcoxon Signed-Rank Test \cite{wilcoxon1992individual} and utilized Cliff's $|\delta|$ effect size \cite{cliff1993dominance} to quantify the extent of the differences.

\noindent\textbf{Continuity.}
Previous work \cite{DBLP:conf/aaai/GhorbaniAZ19} has demonstrated that slight variations in the input samples can confuse the explanation results. If the model response to similar samples varies significantly, not only will it not convince the experts, but it will lead them to doubt the reliability of the underlying predictions. Consequently, Continuity is proposed to describe how continuous and generalizable the explanation function is. In our identified studies, similarity-based metrics, such as \emph{Dice Coefficient} \cite{DBLP:conf/issta/HuWLPWZ023} and \emph{Jaccard Similarity} \cite{DBLP:conf/kbse/LiZZWLXCJ23}, are often utilized to measure the continuity of generated explanations. Higher continuity indicates a better generalizability to new contexts.

\noindent\textbf{Contrastivety.}
Contrastivety aims to describe the discriminativeness of an explanation. Intuitively, the explanation for a particular target or model output should be different from an explanation for another target. \emph{\%Unique} is the most frequent metric, used in two studies \cite{Pyexplainer,DBLP:journals/jss/YangZZXZQ24}. \%Unique measures the percentage of unique explanations generated by each technique. The higher percentage of unique explanations indicates that the explainer can effectively generate a more specific (i.e., less duplicate) explanation to the target instance.

\noindent\textbf{Compactness.}
To avoid increasing the human cognitive load, explanations should be sparse, short and not redundant. As a consequence, compactness is defined to measure the size of generated explanations. Metrics like \emph{Reduction Ratio} and \emph{\#Rules} are the most commonly used, appearing in two and three studies, respectively. For instance, Rabin et al. \cite{DBLP:conf/sigsoft/RabinHA21} employed size reduction ratio of input programs to evaluate the conciseness of generated explanations.

\noindent\textbf{Coherence.}
In many scenarios, even when interacting with the same model, stakeholders who have different intents and expertise may consume explanations for distinct objectives. Hence, assessing to what extent the explanation is consistent with relevant background knowledge, beliefs and general consensus is necessary. The most commonly used coherence metrics are \emph{Alignment-based} (e.g., ROUGE-L, BLEU), \emph{Classification-based} (e.g., Accuracy, Precision, Recall), \emph{Ranking-based} (e.g., Top K-based, MRR), and \emph{Statistical Test}. For example, Fu et al. \cite{VulExplainer} employed Accuracy and Weighted F1-score to evaluate the performance of vulnerability classification. Sun et al. \cite{DBLP:conf/icse/SunXLXZHZ23} adopted ROUGE-1/2/L to measure the usefulness of generated key aspects for alert prediction.

In addition to these popular functional metrics, we also observed a limited number of non-functional metrics. One representative example is \textbf{Efficiency}, which is used in nine studies \cite{FOX,WheCha,DBLP:conf/sigsoft/RabinHA21,DBLP:journals/pacmse/Chen00024}. Most SE tasks, such as code search and code completion, have a high demand for the response efficiency of AI models, i.e., preferring tools/approaches providing actionable results within acceptable time cost, and explanations are no exception. Given that explainability mostly serves as a by-product of model outputs, approaches requiring higher computation overhead are unlikely to be adopted by the audience, even if they achieve promising performance in terms of other aspects.

\begin{myhbox}{\ding{45} $\blacktriangleright$ RQ$_{3}$ - Summary $\blacktriangleleft$}
\begin{itemize}[leftmargin=2em]
\item The analysis indicated a notable scarcity of well-documented and reusable baselines or benchmarks for work on XAI4SE. Approximately 28.7\% of the benchmarks employed in the evaluations of our studied approaches were self-generated, with a significant portion not being publicly accessible or reusable.

\item We noticed that there is no consensus on evaluation strategies for XAI4SE studies, and in many cases, the evaluation is only based on specific properties, such as correctness and coherence, or researchers' subjective intuition of what constitutes a good explanation.

%\item The evaluation metrics predominantly utilized across the studies were categorized based on five types of explanation formats, i.e., numeric, text, visualization, source code, and rule. A significant portion of the reviewed papers employed conventional classification metrics, such as Accuracy and F1-score, to evaluate the performance of their proposed approaches.
\end{itemize}
\end{myhbox}

\section{Discussion}\label{ChaAndOpp}
In this section, we discuss current challenges (Section \ref{Challenges}) and highlight promising opportunities (Section \ref{Opportunities}) for conducting future work on XAI4SE.
\subsection{Challenges}\label{Challenges}
\noindent\textbf{Challenge 1: Lack of Consensus on Explainability in SE.}
One of the major challenges in developing explainable approaches for AI4SE models is the lack of formal consensus on explainability within the field of SE. As shown in the earlier sections, numerous points of view are proposed when trying to articulate explainability for a specific SE task. For instance, to assist software developers in understanding defective commit, some approaches simply highlight the lines of code that the model thinks are defective \cite{DBLP:journals/tse/Wattanakriengkrai22}, while others extract human-understandable rules \cite{Pyexplainer}, or even natural language descriptions \cite{DBLP:conf/icse/MahbubSR23} from the defective code that can serve as actionable and reusable patterns or knowledge. Although this diversity can meet the distinct requirements of audiences with different levels of expertise, it greatly increases the difficulty of establishing a unified framework which provides common ground for researchers to contribute toward the properly defined needs and challenges of the field.

\noindent\textbf{Challenge 2: Trade-off between Performance and Explainability.}
For some real-world tasks, a model with higher accuracy usually offers less explainability. Such \emph{Performance-Explainability Trade-off (PET)} dilemma often results in user hesitation when choosing between black-box models and inherently transparent models. While certain studies \cite{rudin2019stop,rudin2019we} indicate that black-box models performing complex operations do not necessarily result in better performance than simpler ones, it is often the case for advanced SE models built upon immense amounts of structured and unstructured data. This challenge highlights the necessity of flexibly selecting XAI techniques according to various factors, such as the characteristics (e.g., input/output format, model architecture) of different SE tasks, resource availability (e.g., time cost), and consideration of risk (e.g., ethics, legality). For example, explainability may be particularly important to avoid bias in a developer recommendation system, while performance might be prioritized in an AI-aided coding scenario.

%For example, giving precedence to transparent models (e.g., decision trees) when their performance could meet the users' requirements, and transitioning to more complex models when needed.

\noindent\textbf{Challenge 3: Disconnection between Academic Efforts and Industry Needs.}
The deployment of XAI techniques in SE practice, particularly security-critical tasks like vulnerability detection \cite{DBLP:conf/kbse/Cao000B00L024}, necessitates rigorous validation to satisfy not only effectiveness, but also robustness, controllability, and other special concerns. Here, the controllability means that each user should be shown the most complex explanation this user can still grasp, i.e., giving control back to the user. This property is important because users can adapt the explanation to their needs \cite{DBLP:journals/corr/abs-2404-12762}. This proves challenging given the complex and dynamic nature of SDLC. In addition, due to the different intents and expertise of audiences, a single explanation format may not be applicable to everyone. For instance, while visual explanation can be attractive for the layperson since it is natural and intuitive, it could potentially overwhelm domain experts with superfluous information, thereby increasing their cognitive load and rendering the tool counterproductive \cite{DBLP:journals/corr/abs-2309-11960}.

\subsection{Opportunities}\label{Opportunities}
\textbf{Opportunity 1: Application on Underexplored and Complex SE Tasks.} Throughout our study, it was clear that XAI techniques have been widely used in certain SE tasks to support users in decision-making or improve the transparency of AI models. However, the current application of XAI techniques in some SE activities remains relatively sparse. As shown in Figure \ref{RQ1A}, not many studies focus on software requirements \& design and software management. This unveils a substantial opportunity: broadening the application of XAI techniques to these under-explored research topics. We suggest that future work should concentrate on two aspects. First, for emerging SE tasks that have only recently benefited from ML/DL, integrating off-the-shelf XAI techniques into existing research workflows as a component rather than developing end-to-end solutions appears more pragmatic. For example, compared to conventional tools which heavily rely on pre-defined templates or grammars, leveraging the powerful capability of LLMs on code and natural language comprehension to generate formal program specifications \cite{SpecGen}, fix vulnerabilities \cite{DBLP:conf/issta/Xia024}, and analyze developers' sentiment \cite{ting} have shown promising results. By combining them with established techniques \cite{DBLP:conf/nips/MengBAB22}, we can achieve reliable and efficient explanations in a seamless manner. Second, for complex SE tasks that have yet to be fully explored by the research community, such as \emph{Code Search} and \emph{API Recommendation}, it is promising to conduct user survey (e.g., interviewing relevant industrial practitioners) to understand their perceptions.

\noindent\textbf{Opportunity 2: Customizing Task-oriented XAI.}
In the midst of our analysis it became clear that most approaches used to provide explanations for AI-driven SE models are directly inherited from off-the-shelf XAI techniques without any customization. Unfortunately, existing XAI techniques, originally not designed for SE tasks, likely generate suboptimal or even misleading explanations. Given these reasons, we recognize a research opportunity to customize explanation approaches more suitable for SE tasks. In this regard, the integration with domain knowledge appears to be the most promising direction to explore. For example, security experts can construct a vulnerability knowledge base by extracting multi-dimensional information from trusted and public-available intelligence sources such as Snyk\footnote{\url{https://snyk.io/}} and National Vulnerability Database\footnote{\url{https://nvd.nist.gov/}} (NVD). This data- and knowledge-driven strategy fosters not only users' understanding of how a black-box model works, but also their active engagement in its development and evolution.

\noindent\textbf{Opportunity 3: Combination of Various Explanation Formats.}
An obvious trend in our analysis was that different explanation formats are commonly used alone. Given their complementarity (e.g., source code explanations and textual explanation are complementary to each other in an AI-assisted programming support context), we believe there is an opportunity to combine diverse formats of explanation for a more robust and complete decision performance, and improves the likelihood of having at least one explanation that the user understands.

\noindent\textbf{Opportunity 4: Human-in-the-Loop Interaction.}
Exploring a more user-centric approach that provides users with greater agency could lead to results that are orthogonally beneficial to those found using more common techniques. To effectively support human decision-making, there is an escalating need for interactive XAI tools that empower users to actively engage with and explore black-box SE models, thereby facilitating a profound comprehension of the models' mechanisms and their explainability. However, most reviewed works leave aside important aspects pertaining to the XAI tool’s interaction with SE practitioners as an AI assistant. Several studies have highlighted that users had more trust when presented with interactive explanations \cite{DBLP:conf/iva/WeitzSSHA19}. Thus, one possible research interest could be in the application of human-AI dialogue agent. For instance, LLMs such as ChatGPT\footnote{\url{https://chatgpt.com/}} can serve as the main controller or ``brain'' to guide users to clarify their vague intents through multi-round dialogue and return personalized explanations by having access to one or more XAI techniques.

\noindent\textbf{Opportunity 5: Curating High-quality and Multi-dimensional Benchmarks.}
Evaluating the newly proposed approach over baseline techniques on a benchmark is developing into a standard practice in SE research. Nevertheless, existing benchmarks may face issues related to quality, availability, and personalization. First, ground truth information scarcity is still a major issue since the process of data annotation is expertise-intensive and time-consuming for large-scale datasets. This is even more critical in the XAI field, where additional (and commonly multi-modal) annotations are required (e.g., textual descriptions, structured decision-making rules). A potential solution involves promoting cooperation and collaboration between the industry and academia. We have started to see efforts in constructing high-quality benchmarks with annotated explanatory information in other domains, such as the FFA-IR dataset \cite{DBLP:conf/nips/LiCLWZWCLPLZLSV21} for evaluating the explainability on medical report generation. Second, the most adopted evaluation approach in current XAI4SE studies is to resort to the practitioners’ expertise. However, considering the variability in experts’ opinions, this strategy is particularly biased and subjective \cite{10.1145/3583558}. Given that such human feedback is immensely valuable in understanding the strengths and weaknesses of XAI techniques, a clear avenue for improvement is to standardize a protocol for human evaluation of these systems by SE practitioners.

\section{Guidelines for Future Work on XAI4SE}\label{Guideline}
\begin{figure}[t]
  \centering
  \includegraphics[width=.95\linewidth]{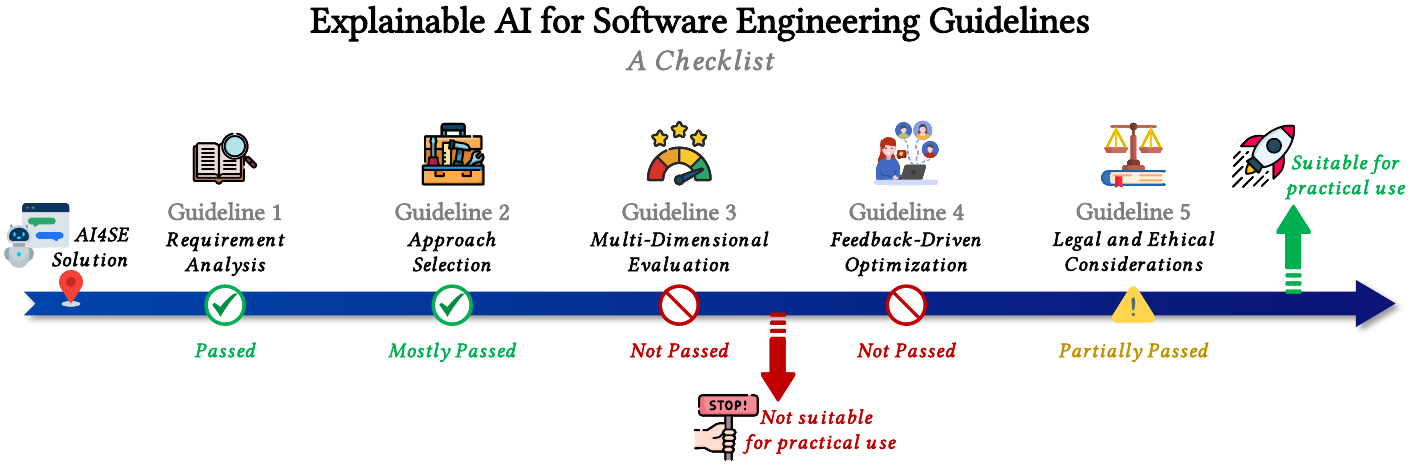}
\caption{Guidelines for applying XAI to SE research.}
\label{Guidelines}
\Description{}
\end{figure}

In this section, we synthesized a checklist with five prescribed steps that should aid in guiding researchers through the process of applying XAI in SE, as illustrated in Figure \ref{Guidelines}.

\noindent\textbf{Guideline 1: Requirement Analysis.}
This first step focuses on clarifying specific needs of distinct stakeholders for a certain SE task. For example, for end-users (e.g., developers, practitioners) which do not have technical knowledge in underlying AI models, the format and context of an explanation should be easily understandable so that they can seamlessly incorporate such evidence in their decision process. This involves (\ding{182}) determining who uses the model outputs, and in what way; (\ding{183}) refining requirements through user surveys; and (\ding{184}) standardizing requirements into document form by integrating the feedback.

\noindent\textbf{Guideline 2: Approach Selection.}
While one of the advantages of out-of-the-box XAI toolkits is that they provide some degree of convenience, it is still important for researchers carefully consider various aspects of existing explanation techniques and determine whether a given technique could be adapted for their task to prevent the oversimplification of the problem, or whether the creation of a new technique should be considered. As such, this step involves (\ding{182}) determining suitable explanation techniques based on task fitness, model compatibility, and stakeholder preference; and (\ding{183}) accounting for the combination of different formats of explanation. Additionally, all steps should be thoroughly documented to support replication.

\noindent\textbf{Guideline 3: Multi-Dimensional Evaluation.}
Explainability is a multi-faceted concept. Thus, this step requires (\ding{182}) carefully choosing metrics, especially quantitative measures beyond accuracy, for the task; (\ding{183}) testing the approach against well-constructed benchmarks; and (\ding{184}) ensuring it has reached optimum capability. Such a multi-dimensional overview could be implemented as a radar chart that comprehensively and concisely conveys the strengths and weaknesses of the explanation approach. Similar to the previous step, researchers should strive to both include the details of their evaluation plan as well as provide rationale for the choices made.

\noindent\textbf{Guideline 4: Feedback-Driven Optimization.}
In addition to quantitative evaluation, it is also crucial to investigate the usability of XAI techniques from a human-centric view. This step demonstrates a meaningful attempt to intuitively understand how the proposed approach would perform in a real-world scenario. Here, researchers should integrate XAI4SE solutions into developers' daily development workflows gradually, and conduct empirical studies, such as interviews, questionnaires, and observation of developers using these techniques in real-world scenarios, to gather rich user experiences. Such feedback helps to identify pain points that can further enhance user satisfaction.

\noindent\textbf{Guideline 5: Legal and Ethical Considerations.}
The final step involves properly evaluating the potential legal and ethical implications before deploying XAI techniques in the wild. Specifically, it is necessary to ensure your data collection process is \emph{compliant}, \emph{privacy-preserving}, and \emph{unbiased} \cite{DBLP:journals/tifs/ZhangZSGCSY24,DBLP:journals/tdsc/ZhangZGMZSC24}. Moreover, it is also important to carefully consider the possible consequences of inaccurate explanations. Therefore, researchers should take appropriate measures, e.g., conducting audits in a regular manner, to minimize any risks of this kind.

\section{Conclusion}\label{Conclusion}
Explainability remains a pivotal area of interest within the SE community, particularly as increasingly advanced AI models rapidly advance the field. This paper conducts a systematic literature review of 108 primary studies on XAI4SE research from top-tier SE and AI conferences and journals. Initially, we formulated a series of research questions aimed at exploring the application of XAI techniques in SE. Our analysis began by highlighting SE tasks that have significantly benefited from XAI, illustrating the tangible contributions of XAI (RQ1). Subsequently, we delved into the variety of XAI techniques applied to SE tasks, examining their unique characteristics and output formats (RQ2). Following this, we investigated the existing benchmarks, including available baselines, prevalent benchmarks, and commonly employed evaluation metrics, to determine their validity and trustworthiness (RQ3).

Despite the significant contributions made to date, this review also uncovers certain limitations and challenges inherent in existing XAI4SE research, offering a set of guidelines that delineate promising avenues for future exploration. It is our aspiration that this SLR equips future SE researchers with the essential knowledge and insights required for innovative applications of XAI.

\bibliographystyle{ACM-Reference-Format}
\bibliography{sample-manuscript}

\end{document}